\documentclass[
  journal=largeone,
  manuscript=article-type,
  year=xxxx,
  volume=xx,
]{cup-journal}

\usepackage{amsmath}
\usepackage[nopatch]{microtype}
\usepackage{booktabs}
\usepackage{longtable}
\usepackage{biblatex}

\title{Limits on Unintended Radio Emission from Geostationary and Geosynchronous Satellites in the SKA-Low Frequency Range}

\author{Tingay, S.J.}
\affiliation{International Centre for Radio Astronomy Research, Curtin University, Bentley, WA 6102, Australia}
\email[S.J. Tingay]{s.tingay@curtin.edu.au}

\author{Hurley-Walker, N.}
\affiliation{International Centre for Radio Astronomy Research, Curtin University, Bentley, WA 6102, Australia}

\author{Ross, K.}
\affiliation{Australian SKA Regional Centre, Curtin University, Bentley, WA 6102, Australia}

\author{Galvin, T.J.}
\affiliation{Australia Telescope National Facility, CSIRO, Space and Astronomy, PO Box 1130, Bentley, WA 6151, Australia}

\author{Morgan, J.}
\affiliation{Australia Telescope National Facility, CSIRO, Space and Astronomy, PO Box 1130, Bentley, WA 6151, Australia}

\author{Venville, B.}
\affiliation{International Centre for Radio Astronomy Research, Curtin University, Bentley, WA 6102, Australia}


\begin{document}

\begin{abstract}
We search data from the GLEAM-X survey, obtained with the Murchison Widefield Array (MWA) in 2020, for the presence of radio frequency interference from distant Earth-orbiting satellites, in the form of unintended emissions similar to those recently seen from objects in Low Earth Orbits (LEO).  Using the GLEAM-X $\delta=1.6^{\circ}$ pointing, which is stationary in azimuth (on the local Meridian) and elevation (near the celestial Equator), the very wide field of view of the MWA maintains custody of a large number of satellites in geostationary and geosynchronous (GEO) orbits in this direction for long periods of time.  We use one night of GLEAM-X data in the 72 - 231 MHz frequency range to form stacked images at the predicted coordinates of up to 162 such satellites, in order to search for unintended radio emission.  In the majority of cases, we reach 4$\sigma$ upper limits of better than 1 mW Equivalent Isotropic Radiated Power (EIRP) in a 30.72 MHz bandwidth (dual polarisation), with the best limits below 10 $\mu$W.  No convincing evidence for unintended emissions at these detection thresholds was found.  This study builds on recent work showing an increasing prevalence of unintended emissions from satellites in LEO.  Any such emission from objects in GEO could be a significant contributor to radio frequency interference experienced by the low frequency Square Kilometre Array and warrants monitoring.  The current study forms a baseline for comparisons to future monitoring.
\end{abstract}

\section{Introduction}
The increasing number of active satellites in Earth orbit has been shown to have a negative impact on radio astronomy.  Radiation unintentionally emitted outside frequency bands allocated to satellite communications has been recently detected at levels comparable to the strongest cosmic radio sources in the sky at low radio frequencies \citep{2025A&A...698A.244Z, 2025A&A...699A.307G,2024arXiv241214483G,2024A&A...689L..10B,2023A&A...678L...6G,2023A&A...676A..75D}.  While unintended radiation of this nature is currently not subject to regulation by the International Telecommunications Union (ITU), nonetheless the  unintended emission constitutes Radio Frequency Interference (RFI) for radio telescopes \citep{2024AcAau.225.1019M}.

The majority of this impact has generally been discussed in relation to the greatly increased use of the Low Earth Orbit (LEO) regime for global communications networks, supported by large scale constellations of satellites, but other orbital regimes are also relevant.  Geostationary and Geosynchronous Earth Orbits (GEO) have long been used for communications systems and Earth observation purposes (e.g. weather observations), due to the convenient property that satellites in GEO can maintain their positions over fixed positions on the Earth's surface, or within a narrow range of longitudes.  At a distance of 35,785 km above the Earth's surface, if a satellite has an orbit around the Earth's Equator, it will appear to remain at a fixed azimuth and elevation for an observer on the Earth's surface and be classified as geostationary.  If a satellite at this distance has a non-zero inclination to the Equator, a ``figure 8'' motion is traced on the sky for an observer on the ground, within an azimuth and elevation range, and the object is classified as geosynchronous.  Generally when geostationary/synchronous satellites are de-commissioned, they are moved into slightly higher orbits, known as Graveyard Orbits, causing the satellites to slowly drift to the west and gradually become more inclined to the Equator due to various gravitational perturbations.  Operational satellites in the GEO regime are required to maintain their positions within well-defined spatial boxes. As with LEO, the GEO regime is increasingly congested as utilisation increases.  From an Earth-based observer point of view at the mid-latitudes, approximately equal numbers of LEO and GEO satellites will be above the local horizon at any given point in time, although the numbers in LEO are increasing at a faster rate than for GEO.  As is the case for satellites in LEO, astronomy is affected by intentional transmissions in allocated communications bands (between 2 and 20 GHz) by satellites in GEO\footnote{https://science.nrao.edu/facilities/vla/docs/manuals/obsguide/rfi\#autotoc-item-autotoc-4}.

While any RFI from satellites in LEO or GEO can pose a problem for radio astronomy, the effects in the two regimes may be seen as quite different.  An object in LEO travels at approximately 8 km/s, producing an angular speed across the sky for an observer on the Earth of approximately 1 degree/s near zenith, much faster than the sidereal rate used to track astronomical radio sources.  This means that the response of a radio interferometer to radiation from an object in LEO will be greatly reduced (but not completely eliminated) by smearing effects \citep{2022AdSpR..70..812P}.  The smearing will depend on the direction of motion on the sky and the angular speed of the object, as well as the length and orientation of the baselines between antenna pairs of the interferometer and the observation frequency.  In addition, objects in LEO are in the near field of modern-sized interferometers, leading to additional baseline-dependent smearing effects \citep{2025PASA...42...10D,2023PASA...40...56P}.  The interferometric response is, thus, very complicated, with a complex impact on high precision experiments such as the detection of the redshifted neutral hydrogen signal from the Epoch of Reionisation \citep{2020MNRAS.498..265W}.  The danger for high precision experiments is that the RFI contaminants are reduced through smearing but not fully removed, pushed below the noise level such that they only manifest after very long integrations.  At that point, identifying and excising the RFI may be difficult, although emerging techniques are seeking to address this issue \citep{2023MNRAS.524.3231F}

For objects in GEO, the angular speeds at which the objects are seen to move from the surface of the Earth are much slower.  This has several consequences.  The objects are moving at rates much closer to the sidereal rate of $\sim$15 arcseconds per second with respect to the astronomical sky, and thus smearing effects will be significantly reduced.  While this may make the presence of RFI in interferometric data more obvious, the received power from an object in GEO may be far lower than from an object in LEO, due to the inverse square law.  An object in LEO that produces a 10 Jy signal at a radio telescope would produce a signal of approximately 1 mJy if placed at GEO.  Thus, even though the effect of smearing is less, the greatly reduced received power may render signals difficult to detect.  Finally, because of the slow angular speeds, GEO objects can remain in the field of view of a telescope for long periods of time.  This is especially true of modern radio telescopes with very wide fields of view.  However, the portion of the sky affected will generally be mostly concentrated around the celestial Equator, not across the entire sky.

Following previous work that has examined the impact of satellite constellations in LEO, at frequencies relevant to the future low frequency Square Kilometre Array (SKA-Low: \cite{5136190}), we consider the points above and undertake a study to search for unintended radio emissions from satellites in GEO.  Given the brief description of the relevant orbital dynamics provided above, this search represents a snapshot in time, as satellites in Graveyard Orbits move in and out of view, as new satellites appear, and as old satellites are de-commissioned.  We utilise data already acquired for the GLEAM-X survey \citep{2025PASA...42..129H}, collected using the Murchison Widefield Array (MWA: \cite{2018PASA...35...33W,2013PASA...30....7T}).  The GLEAM-X data products we use were the result of specific processing to support searches for transient radio sources, which have yielded several high impact discoveries \citep{2024ApJ...976L..21H,2025ApJ...981..143M}.  For the purposes of studying GEO satellites, these data are extremely convenient as the GLEAM-X survey was configured as a sequence of ``drift scans'', whereby the MWA field of view was placed on the local meridian at different declinations, allowing the sky to pass through the field of view over the course of a night.  Observing frequencies were switched over the course of a night to produce large-scale, multi-frequency drift scan data over wide fields of view.  We choose the Equatorial pointing, which encompasses a large number of satellites at GEO distances that we can examine.  The data we use cover the 72 - 231 MHz frequency range, which is a significant fraction of the planned SKA-Low frequency range of 50 - 350 MHz.  Within these frequency ranges, several relatively narrow frequency bands have various levels of protection for radio astronomy under the ITU regulations; see \cite{2023A&A...678L...6G} for a summary.

In \S 2 we describe the details of the data we have used and the data processing applied.  In \S 3 we present and discuss the results of the data processing.

\section{The Data and Data Analysis}

\subsection{The Data and Imaging/Calibration Pre-processing}

We use data obtained by the MWA for the GLEAM-X survey, already calibrated and imaged for the purpose of searching for transient radio sources.  The GLEAM-X survey observations were structured as $\sim$100 drift scans, whereby the MWA beam was pointed at the local meridian and a fixed declination and over the course of a night the sky drifted through the beam.  During a night, the MWA's 30.72 MHz contiguous bandwidth was switched between five centre frequencies so that five drift scans were produced, covering a frequency range of 72 to 231 MHz (centre frequencies of 87.675, 118.395, 154.235, 184.955, and 215.675 MHz, hereafter referred to 88, 118, 154, 185, and 216 MHz).  The data were collected in 120 second blocks, each identifed as an MWA Observation ID (ObsID).  An ObsID contains data at only a single frequency band.

For each ObsID, data were delivered to us already calibrated and imaged using WSClean, \citep{2014MNRAS.444..606O} providing images for each MWA polarisation (XX and YY) that were averaged over the full 30.72 MHz bandwidth and at a cadence of 4 seconds.  Further, without correction for the primary beam, the XX and YY images {\bf had been} added together to produce pseudo Stokes I images.  The images had been cleaned and the output was stored in an HDF5 format that contains header information and metadata, clean models, and residual images.  Each HDF5 container has a size of approximately 300 MB. Our starting points for the analysis in this paper are the HDF5 files.  We use data collected during the night of 2020-10-10 (ObsIDs ranging from 1286363152 to 1286398432) for the $\delta=1.6^{\circ}$ drift scan, which covers the Equatorial region and thus the maximum number of GEO objects.  The data consist of 254 HDF5 files and approximately 74 GB data volume.  Table \ref{tab1} contains relevant imaging parameters at each frequency.  A more detailed description of the data processing steps leading to the HDF5 formatted data products is given in \cite{2024PASA...41...54R}.  For a description of the HDF5 format itself, see Appendix 1 of \cite{2018MNRAS.473.2965M}.

All original data are discoverable at the MWA ASVO archive portal\footnote{https://asvo.mwatelescope.org/}, where they can be downloaded in raw form (original visibilities, not the processed form of data used here).

\begin{table}[hbt!]
\caption{Imaging parameters}
\label{tab1}
\begin{tabular}{cccccc}
\toprule
\headrow Centre Frequency & Frequency Range&Integration time  & Number of pixels & Pixel size & Image size \\
\headrow (MHz)&(MHz)& (s) &&(arcminutes)&($^{\circ}\times^{\circ}$) \\
87.675&72.315 $-$ 103.035 &4&2400$\times$2400&1.04&41.7$\times$41.7 \\
118.395&103.035 $-$ 133.755&4&2400$\times$2400&0.77&31.0$\times$31.0 \\
154.235&133.755 $-$ 169.595&4&2400$\times$2400&0.56&23.8$\times$23.8 \\
184.955&169.595 $-$ 200.315&4&2400$\times$2400&0.50&19.9$\times$19.9 \\
215.675&200.315 $-$ 231.035&4&2400$\times$2400&0.43&17.0$\times$17.0 \\
\end{tabular}
\end{table}

\subsection{Data Processing and Image Stacking}

Using recent measurements of unintended emissions from LEO satellites as a guide \citep{2023A&A...678L...6G}, in the tens to hundreds of Jy range, comparable emitted powers from objects in GEO would be in the mJy regime.  Signals of this level would not be included in the clean models in the HDF5 containers.  Also, at these levels, RFI flagging applied during initial calibration steps will not excise these signals.  So, we take the approach of utilising the residual images, with image stacking to decrease the noise and increase the sensitivity of the stacked images to the point that mJy-level signal detection is plausible.  Implicit in this approach is the assumption that any unintended emission fully occupies the 30.72 MHz bandwidth of the observation, as well as the full temporal period of the images being stacked.  As unintended emissions from objects in LEO have been observed to be broadband in nature and constant in time, this is a reasonable assumption.  In the 4 second images, the motion of the sky across the field of view (manifested as the drift of a GEO satellite across the image) corresponds to approximately 1 arcminute, approximately equal to one pixel in the images.  Thus, minimal loss of signal due to smearing will be experienced.

All steps described henceforth were implemented in a Python script\footnote{Available from https://github.com/steven-tingay/RFI-from-GEO}.  In order to stack images over time, centred on the location of an individual satellite, information on the predicted location of the satellite is required.  We obtained this information via an API query to the space-track.org database of space situational awareness information.  The query requested return of Two Line Elements (TLEs) for all tracked objects that have a semi-major axis within 5\% of the geostationary distance and an ellipticity $<$ 0.01, with TLE generation dates contained in a one week window centred on 2020-10-10.  TLEs for 1396 unique objects were returned.

From the HDF5 containers for ObsIDs at a given frequency, the residual images (FITS format) were extracted, along with their time stamps.  From the time stamps, predicted Right Ascension and Declination values for a satellite, as observed from the MWA location, were generated using the Skyfield python module \citep{2019ascl.soft07024R}\footnote{https://rhodesmill.org/skyfield/}.  Using the World Coordinate System (WCS) extracted from the residual image FITS header information, if the satellite was found to be within the imaged MWA field of view, a 100$\times$100 pixel cutout of the image, centred on the predicted satellite coordinates, was generated (positions needed to be more than 50 pixels from an edge of the image).  

Before being added to a stack, each individual cutout image was corrected for the response of the MWA primary beam\footnote{https://github.com/MWATelescope/mwa\_pb}.  As the residual images are psuedo Stokes I, simply the addition of the orthogonal XX and YY polarisation images, the beam model used to correct the cutouts was constructed from the average of the XX and YY beam models, generated for the appropriate frequency (centre frequency of each 30.72 MHz band) and pointing direction.  The reciprocal of the beam response in the direction of the object was used to multiply all pixels in the cutout, before adding the cutout to the stack for that object.  After all residual images at all times and in all relevant ObsIDs were processed, the stack was divided by the number of cutouts that contributed to the stack.

It should be noted that the stacks are produced in pixel space and the individual cutouts are not rotated and re-sampled, according to the instantaneous direction of motion of the satellite under analysis, before adding to a stack.  When tracking a satellite from horizon to horizon, such rotations and re-sampling are required \citep{2022AdSpR..70..812P}.  For geostationary and geosynchronous satellites over a limited field of view near the celestial equator, however, the rotations and smearing have less impact.  We found that for all 162 satellites ultimately under consideration, all signals were smeared over areas less than two pixels in width, leading to a small maximum ($<$10\%) reduction in peak intensity due to smearing.  This has been disregarded in the subsequent analysis, in favour of not introducing additional complexity into the data processing for minimal gains.

In order to verify that the translation from Right Ascension and Declination to pixel coordinates via the WCS, and the stacking of cutout images, was working correctly, we performed the same set of steps but for two known strong extragalactic radio sources, 3C33 and 3C118.  In this case, all processing steps were identical, but as strong radio sources their signals were captured in the clean models, so we used the combined CLEAN model plus residual image rather than just the residual image.  Although these astronomical sources do not move in celestial coordinates as objects in GEO do, the drift of the sources through the images partially mimics the behaviour of objects in GEO (a geostationary object drifts through celestial coordinates at the sidereal rate).  In both cases, and at all frequencies, the stacked cutouts of 3C33 and 3C118 produced the expected stacked signal at the expected centre pixel.

As the field of view of the MWA is frequency dependent, a larger number of objects was found to be present in the images at the lowest frequency, where the field of view is largest.  We found that 162 objects were present for at least some of the time in the field of view at 88 MHz, with 119 at 118 MHz, 93 at 154 MHz, 76 at 185 MHz, and 63 at 216 MHz.  Thus, a total of 513 stacks were generated across all frequencies.

\subsection{Identification of Candidate Detections}

The RMS of the 100$\times$100 pixel image stacks was measured, to determine detection thresholds.  Further, since the TLE information for satellites in GEO can have an up to 40 km position error \citep{Racelis2018-pj}, corresponding to approximately 4 arcminutes, candidate detections were thus considered in a region of this size, centred on the predicted position.  As the pixel scale is frequency dependent, this region corresponded to $\pm$4 pixels in $x$ and $y$ directions at 88 MHz, $\pm$5 pixels at 118 MHz, $\pm$7 pixels at 154 MHz, $\pm$8 pixels at 185 MHz, and $\pm$9 pixels at 216 MHz.

Assuming Gaussian statistics, indicative detection thresholds can be explored.  First, however, it should be recognised that neighboring pixels in the stacks are not independent (as extensively discussed by \cite{2018PASA...35...11H} in the context of fitting source models to interferometric images), with the number of independent samples in a stack approximately equal to the number of pixels divided by the number of pixels per synthesied beam, which is approximately 10 for our data.  For example, the error region at 88 MHz of 8$\times$8 pixels yields approximately six independent samples.  Adopting a 3$\sigma$ detection threshold would predict approximately one detection per stack due to thermal noise fluctuations, with approximately 0.008 detections per error region per stack at this frequency.  Across the 162 stacks for all objects, one would thus expect approximately one false positive due to noise fluctuations.  The same calculation at our highest frequency of 216 MHz (18$\times$18 pixels and 62 stacks) predicts approximately three false positives.  Adopting 4$\sigma$ as the detection threshold reduces the expected false positive rates to below unity at all frequencies.  Thus, a 4$\sigma$ detection threshold may be reasonable for the stacks at each frequency.  We also consider detections above a 3$\sigma$ threshold, looking for detections at the same location across multiple frequencies at this lower level of significance.  However, it also needs to be noted that, for these two tests, the data are not likely to be perfectly Gaussian distributed in reality, making this type of analysis an approximation.

Therefore, while we undertake threshold testing of possible detections at 3$\sigma$ (across multiple frequencies) and 4$\sigma$ (single frequencies), we also examine other detection metrics.  We use a simple metric that asks if there is an excess in the incidence of the brightest pixel in a stack being in the TLE error region.  For example, for a 100$\times$100 pixel stack with an error region of 8$\times$8 pixels, the chance that random fluctuations produce the brightest pixel in the error region is approximately 0.006.  Across 162 stacks, approximately one detection in the error region may be expected.  An excess beyond a given number of detections in the error region can be ascribed a probability via binomial statistics.  In python we use \rm{scipy.stats.binom.sf($x,n,p$)} to assess the probability of the number of detections greater than some value, $x$, given a number of trials, $n$, and the probability per trial, $p$.  For example, for the $n=162$ trials at 88 MHz, with $p=0.006$, a value of $x\geq3$ has a probability of 0.04.  Thus, detection rates of three or above per 162 stacks can be ruled out as due to random fluctuations with 95\% confidence.  The test does not indicate, however, which individual detections may be identified with satellites in the case of an excess.  The test is robust to the correlation between pixels noted above, but a caveat on the application of this test to this particular dataset, is that the stacks may not, strictly speaking, be fully independent.  For two satellites in our field of view that have similar trajectories across the sky, image information will be common to both stacks and therefore not be fully independent.  Also, it is worth noting that it is possible that real but weak astronomical radio sources appear in the stacks, at the higher sensitivity levels the stacks achieve.

We also run these analyses after multiplying the pixel values in the stacks by -1.  For detections generated by random fluctuations, there should be as many negative fluctuations as positive fluctuations.  Thus, the statistical results in all three tests should be consistent if detections are due to random fluctuations.

For completeness, we also ran full processing of the satellite list stacking the CLEAN model plus residuals, rather than just using the residual images.  The results were barely discernible from each other, which indicates that the temporal occupancy of CLEAN model components in our cutout regions is very low.  This is understandable as the cutouts are $\sim$10 arcminutes on a side, which represents less than one minute in time at the Equator.  We thus consider the residuals-only processing for further discussion.

\section{Results and Discussion}

\begin{figure*}
\centering
\includegraphics[width=0.45\linewidth]{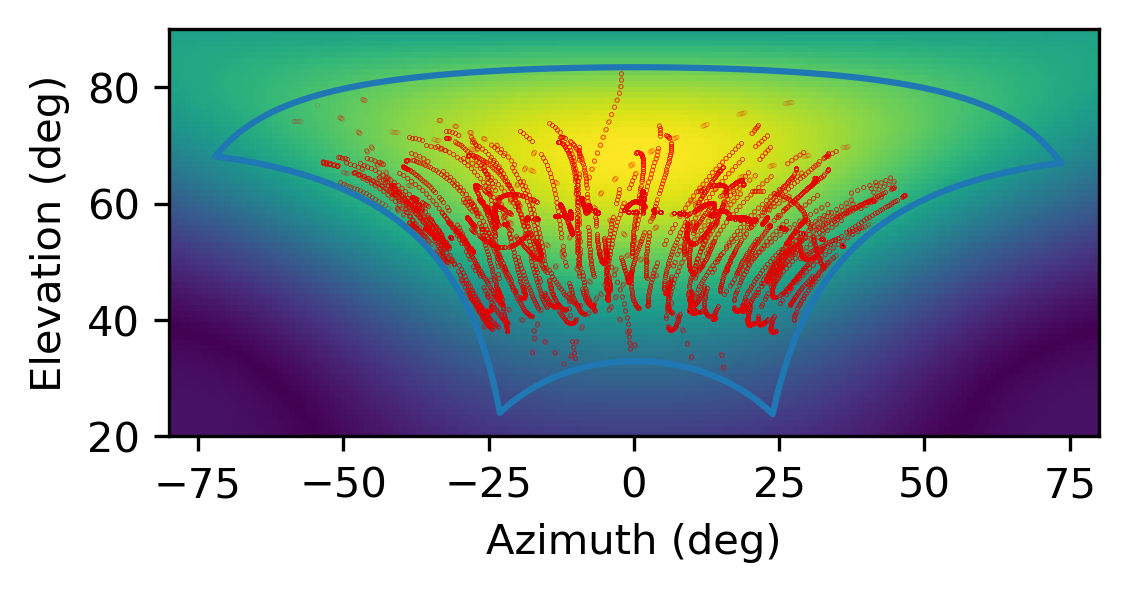}
\includegraphics[width=0.45\linewidth]{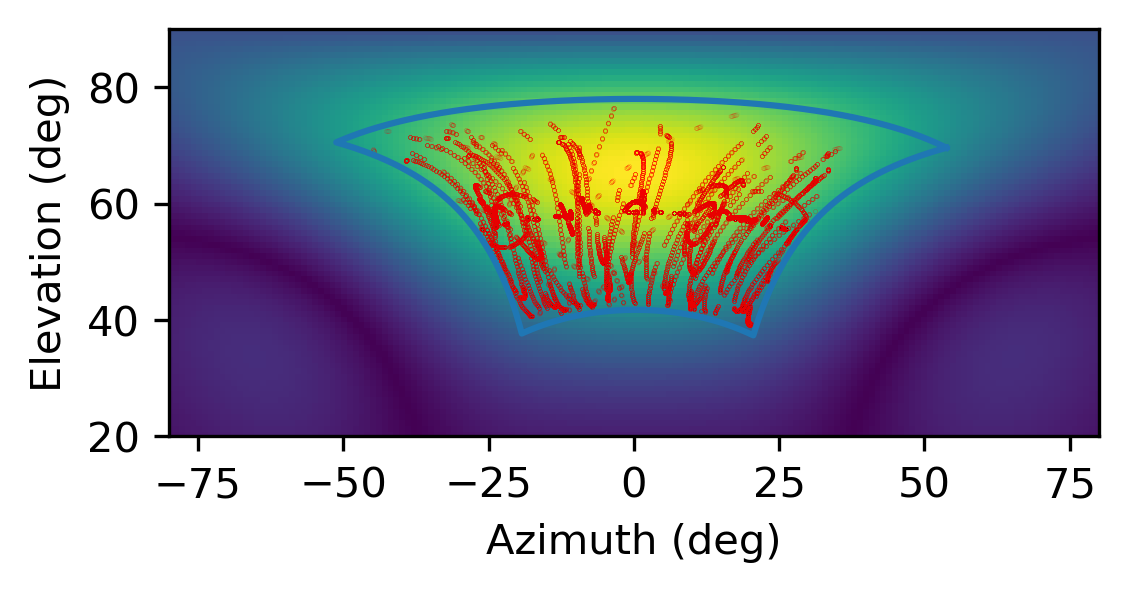}
\includegraphics[width=0.45\linewidth]{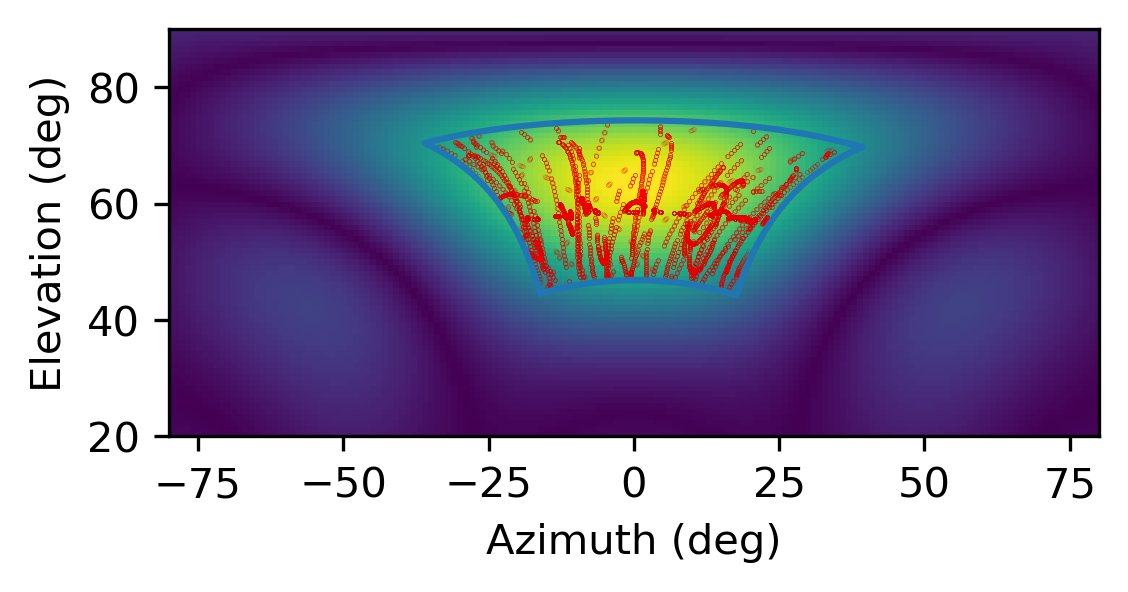}
\includegraphics[width=0.45\linewidth]{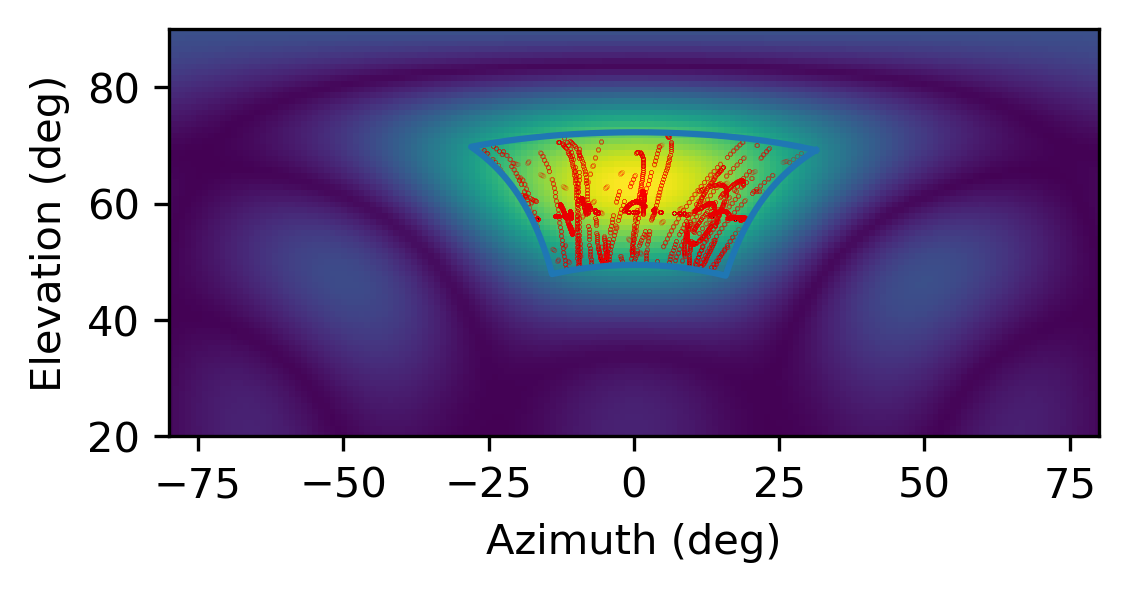}
\includegraphics[width=0.45\linewidth]{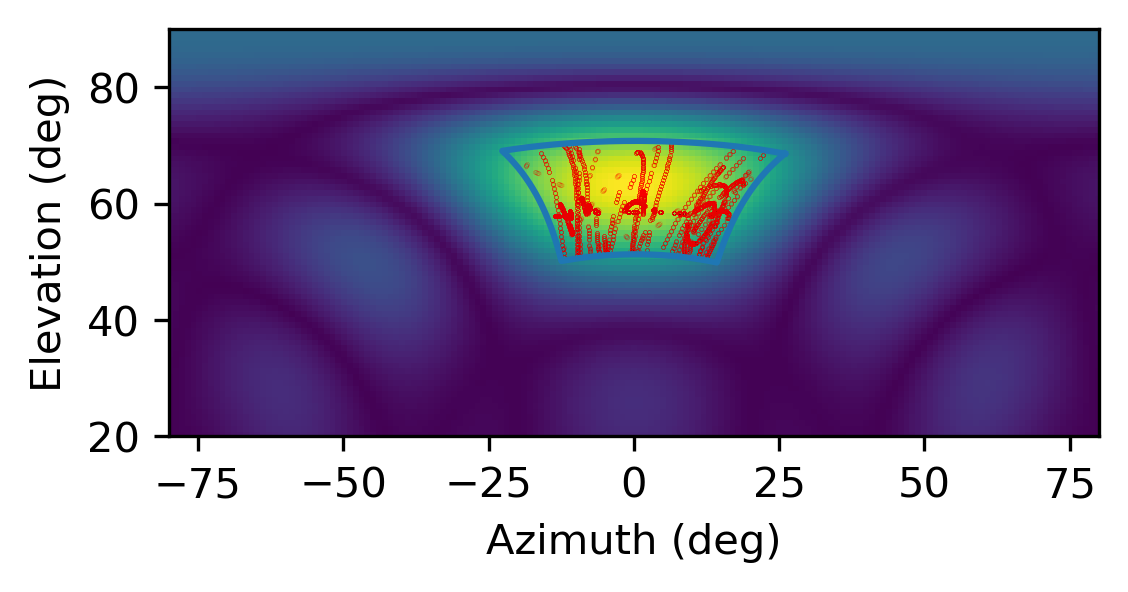}
\caption{MWA imaged field of view in azimuth and elevation coordinates (blue line), with the MWA beam response as a colour grayscale.  Red points are the coordinates of individual satellites at individual times.  Frequencies are 88 MHz to 216 MHz, from top left to bottom.}
\label{all_fields}
\end{figure*}

Figure \ref{all_fields} shows the trajectories of all stacked satellites in the field of view at each frequency.  Table \ref{sum_tab} below summarises the results of our statistical tests, including the 4$\sigma$ detections and the results of the test for excess brightest pixels in the TLE error region (for both the stacks and negated stacks).

In the Appendix, Tables \ref{tab2}, \ref{tab3}, \ref{tab4}, \ref{tab5}, and \ref{tab6} list the parameters for each satellite present in the MWA field of view at each frequency.

\begin{table}[hbt!]
\begin{tabular}{cccccc}
\toprule
\headrow Frequency (MHz) \# & 4$\sigma$ detections  & $N_{ER+}$ & $P_{ER+}$&$N_{ER-}$&$P_{ER-}$  \\
88&43432/4.1/137/36726/0.8& 3/162&0.04&2/162&0.09 \\
118&15236/4.5/41/38090/0.25&1/119&0.33&2/119&0.12 \\
154&27683/4.2/11/36953/0.08&1/93&0.54&2/93&0.28 \\
185&&2/76&0.31&0/76&0.86 \\
216&&4/63&0.05&2/63&0.33 \\
\end{tabular}
\caption{Summary of statistical test results.  The 4$\sigma$ detection column gives the NORAD ID number, detection signal-to-noise, peak intensity of detection (mJy/beam), mean distance (km), and Equivalent Isotropic Radiated Power (EIRP; W) for each detection.  $N_{ER}$ is the number of maxima that occur in the error region and $P_{ER}$ is the probability of that number occurring by chance, as described in the text (with the $+$ and $-$ indicating the stacks and the negated stacks, respectively.} \label{sum_tab}
\end{table}

\subsection{Individual Candidates ($>4\sigma$ Events Within Error Regions)}

At 88 MHz, 1/162 objects exceeded the 4$\sigma$ threshold, NORAD ID 43432 (Table \ref{sum_tab}).    This object is listed as COSMOS 2526, launched by the Commonwealth of Independent States from the Tyuratam Missile and Space Complex on April 18, 2018; it is in a geostationary orbit.  While definitive information is difficult to obtain, the following source claims the satellite to be a communications satellite for the Russian Ministry of Defence (https://weebau.com/satcosmos/2/2526.htm).  At this frequency, the test for excess detections in the error region gave a result that is borderline significant, with a probability due to noise fluctuations of 0.05.

At 118 MHz, 1/119 objects exceeded the 4$\sigma$ threshold, NORAD ID 15236 (Table \ref{sum_tab}).  This object is listed as LEASAT 2, launched by the United States from the Air Force Eastern Test Range on August 30, 1984 via the Space Shuttle; it was one of a series of communications satellites owned by Hughes Aircraft Company and leased to the US military.  A detailed description of the LEASAT satellites can be found here: https://en.wikipedia.org/wiki/Syncom.

At 154 MHz, 1/93 objects exceeded the 4$\sigma$ threshold, NORAD ID 27683 (Table \ref{sum_tab}).  This object is INTELSAT 907, one of the series of Intelsat satellites deployed to support communications over North and South America, Western Europe, and Africa; it was launched from French Guiana on February 15, 2003 (https://www.n2yo.com/satellite/?s=27683).

At 185 MHz, 0/76 objects exceeded the 4$\sigma$ threshold.

At 216 MHz, 0/63 objects exceeded the 4$\sigma$ threshold.  At this frequency, our test for excess detections in the TLE error region yielded four detections out of the 63 stacks, which is borderline significant, with a probability due to noise fluctuations of 0.05.

\begin{figure*}
\centering
\includegraphics[width=0.45\linewidth]{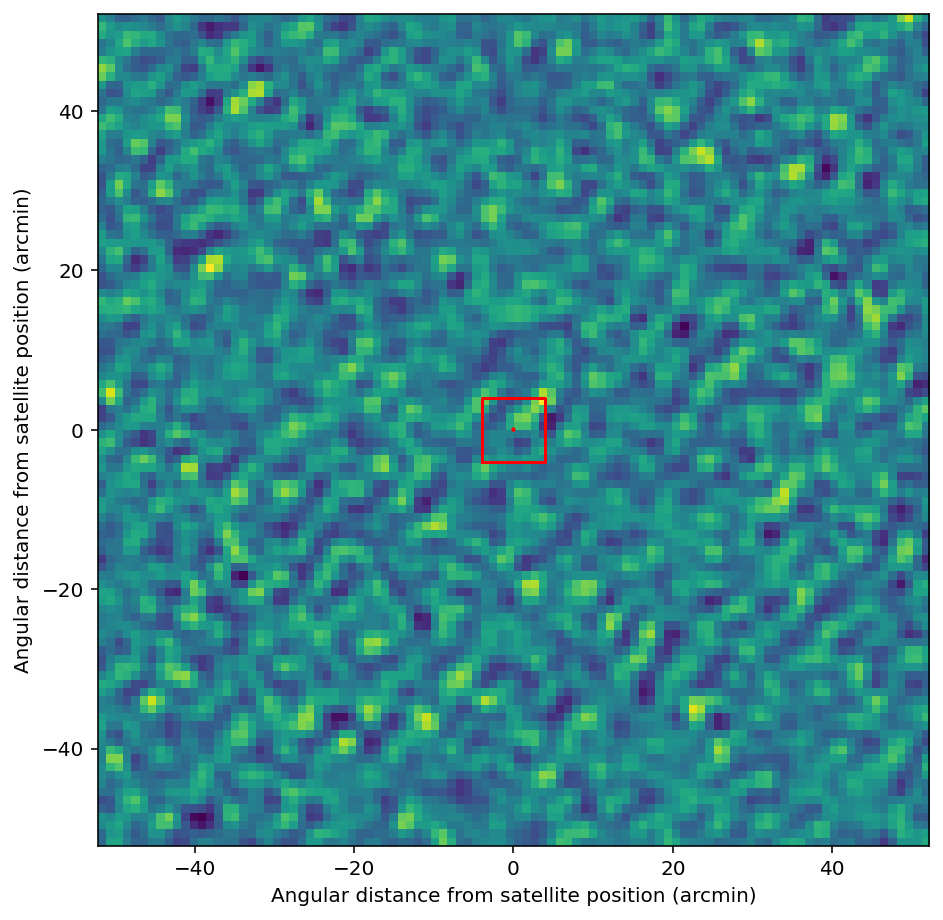}
\includegraphics[width=0.45\linewidth]{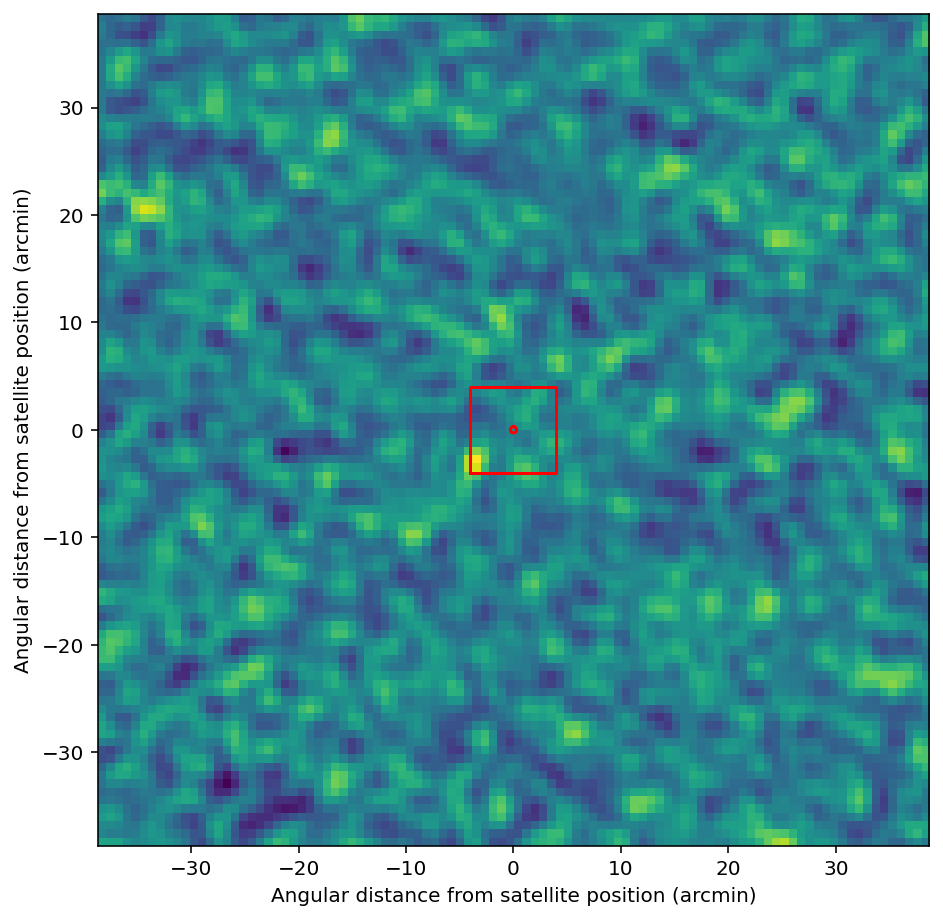}
\includegraphics[width=0.45\linewidth]{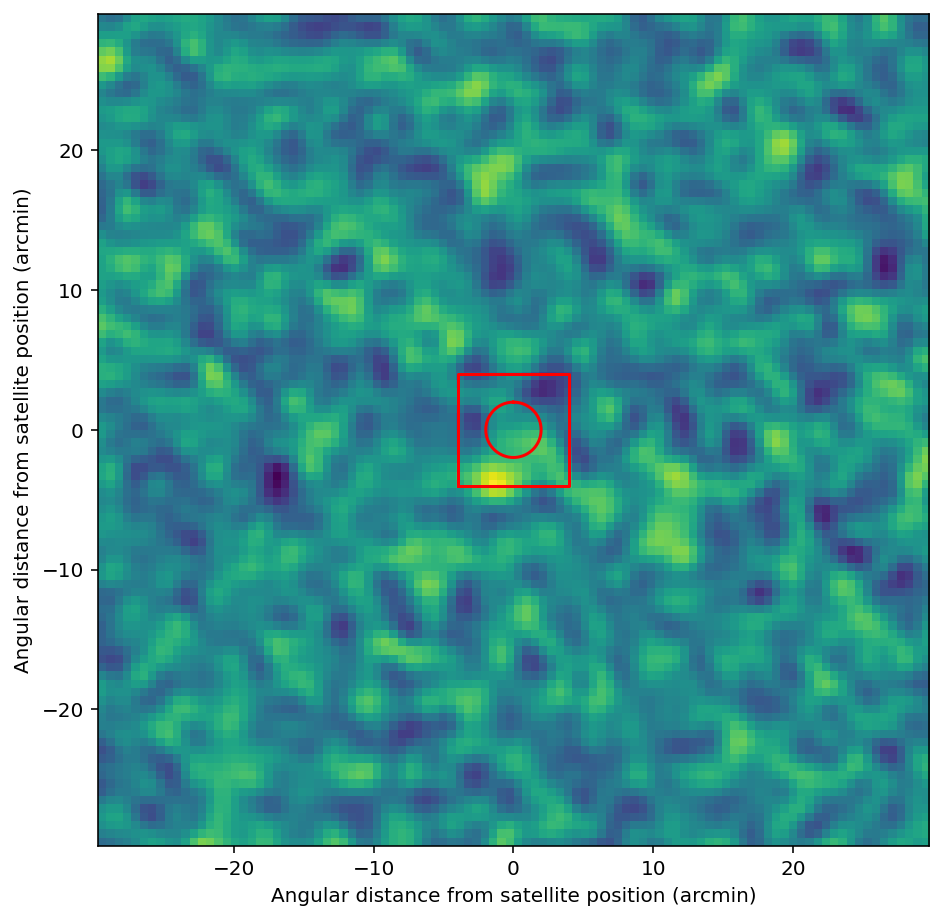}
\caption{The detection stacks for NORAD 43432 at 88 MHz (top left), 15236 at 118 MHz (top right), and 27683 at 154 MHz (bottom). Shown in red are the 4 arcminute error region adopted in the search process and the specific error regions estimated for the individual satellites (shown using the larger of the RA and DEC region values in Table \ref{err-reg} as a circular error region.)}
\label{all_det}
\end{figure*}

Figure \ref{all_det} shows the stacks for NORAD numbers 43432, 15236, and 27683.

While we can adopt a statistical error region for the purposes of our detection algorithms, it is possible to look in detail at the three candidates these algorithms produce, using error region estimates specific to these objects.  If we retrieve TLE data from space-track.org for the three candidate objects, for a two week period centred on 2020 October 10, we can assess the spread in positional predictions from these TLEs, for a particular point in time (2020-10-10:00:00:00 UTC).  This allows us to estimate specific error regions for each object, that we can then compare to the images in Figure \ref{all_det}.  Table \ref{err-reg} describes the resulting estimates.

\begin{table}[hbt!]
\begin{tabular}{cccc}
\toprule
\headrow NORAD ID \# & \# TLEs  & RA RMS (arcmin) & DEC RMS (arcmin)  \\
15236&25&0.07&0.03 \\
27683&37&0.29&0.05 \\
43432&36&1.98&0.16 \\
\end{tabular}
\caption{The Right Ascension and Declination spreads (RMS) calculated from TLEs available in a two week period for the three candidate detections at a reference time of 2020-10-10:00:00:00 UTC} \label{err-reg}
\end{table}

Table \ref{err-reg} shows that the error region estimates for the three candidate objects are far smaller than the 4 arcminute $\times$ 4 arcminute regions assumed above, based on a statistical 40 km positional error.  In particular, when compared to the images in Figure \ref{all_det}, it can be seen that the 4$\sigma$ signals lie outside these smaller error regions.  This is strong evidence that the signals represent statistical fluctuations rather than any origin from these satellites.

The spread for NORAD 43432 is significantly larger than for the other two objects, as 43432 is an active operational satellite that maneuvers often, such that any given TLE will be incorrect.  The other two objects are in graveyard orbits, such that the objects gradually drift, meaning TLEs will be accurate for longer.

\subsection{Further Statistical Tests}

Across all five frequency bands and 513 stacks, three candidate detections above our 4$\sigma$ threshold are apparent, one each at 88, 118, and 154 MHz.  At each frequency a different satellite was identified.  This result is inconsistent with the expected false positive rates calculated above, which are set to be below unity.  However, we note the caveats we have made when it comes to Gaussian statistics.  Thus, we explore further tests.

We note that the smaller field-of-view at higher frequencies reduces the number of opportunities for detection across multiple frequencies.  Looking at the three detections listed above in individual frequency bands, COSMOS 2526 (NORAD 43432) was detected at 88 MHz and was in the field-of-view at 118 MHz (peak intensity of 17 mJy/beam and signal-to-noise of 2.8), but was not in the field-of-view at 154 MHz, 185 MHz, or 216 MHz.  

LEASAT 2 (NORAD 15236) was detected at 118 MHz and was present in the field-of-view at 88 MHz (peak intensity of 57 mJy/beam and signal-to-noise of 1.6), 154 MHz (peak intensity of 18 mJy/beam and signal-to-noise of 2.9), and 185 MHz (peak intensity of 40 mJy/beam and signal-to-noise of 2.6).  LEASAT 2 was not in the field-of-view at 216 MHz.

INTELSAT 907 (NORAD 27683) was detected at 154 MHz and was in the field-of-view at 88 MHz (peak intensity of 87 mJy/beam and signal-to-noise of 3.5), 118 MHz (peak intensity of 10 mJy/beam and signal-to-noise of 2.4), 185 MHz (peak intensity of 6 mJy/beam and signal-to-noise of 2.1), and 216 MHz (peak intensity of 53 mJy/beam and signal-to-noise of 1.7).  INTELSAT 907 was therefore in the field-of-view at all frequencies, but only detected at one frequency.

When we lower our thresholds and test for instances of 3$\sigma$ detections in multiple frequency bands at the same location, we find five instances of $>3\sigma$ detections within the TLE error region common to two frequency bands, but no instances where these detections are coincident in position within the error region.

When running all tests after negating the pixel values in the stacks we find 4$\sigma$ detections at 88 MHz (two), 185 MHz (one), and 216 MHz (one), similar to the non-negated results in terms of detections.  The instances of maxima in the error regions are listed in Table \ref{sum_tab} and are also consistent with the non-negated results.  Finally, in the negated images, we find six instances of $>3\sigma$ detections in the TLE error region with no coincident positions, again consistent with the result above for the non-negated images.

Considering all of the above, the caveats on the use of Gaussian statistics and the consistency of the statistical results between the stacks and the negated stacks, there is no significant evidence for detection of emission from any of these objects.  While we show the three 4$\sigma$ detections in Figure \ref{all_det}, this is mainly to give the reader an idea of the quality of the stacked images.

\begin{figure*}
\includegraphics[width=0.45\linewidth]{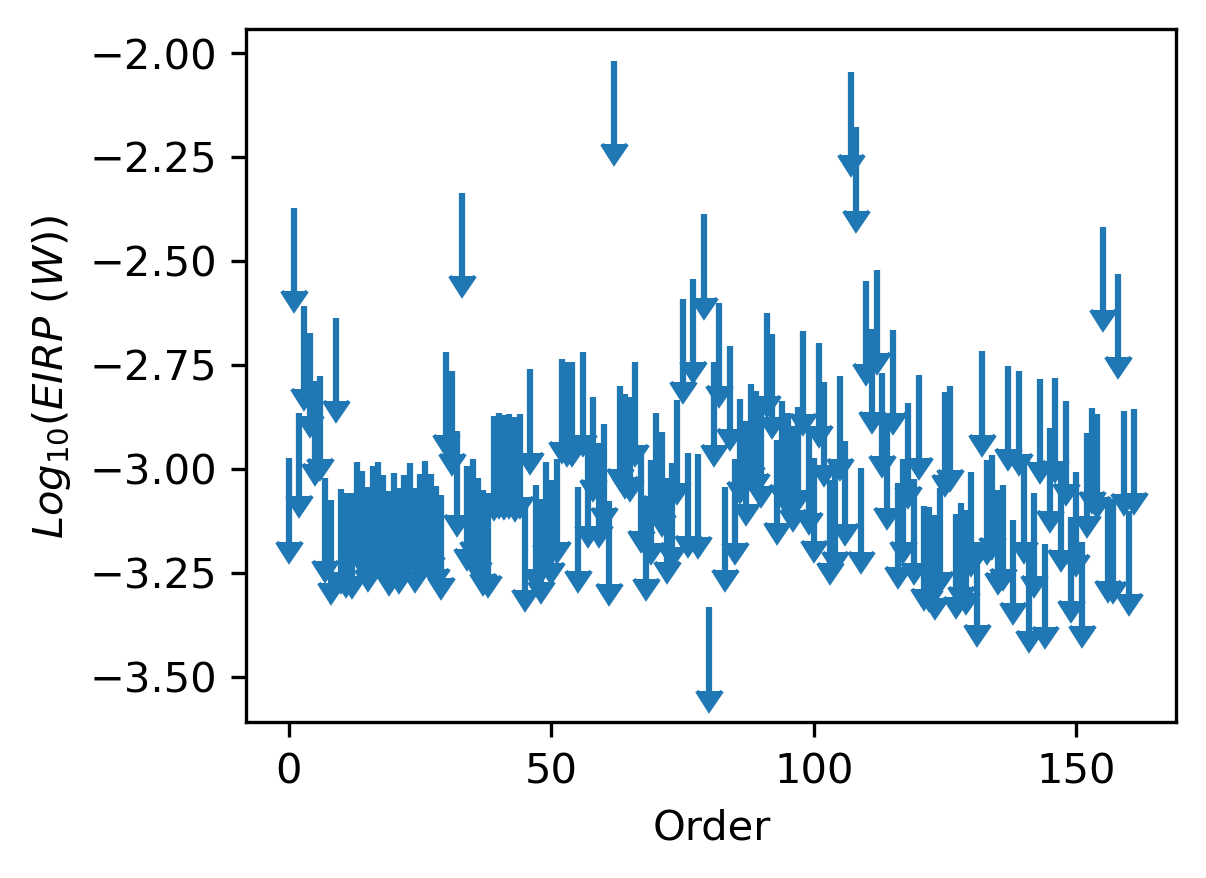}
\includegraphics[width=0.45\linewidth]{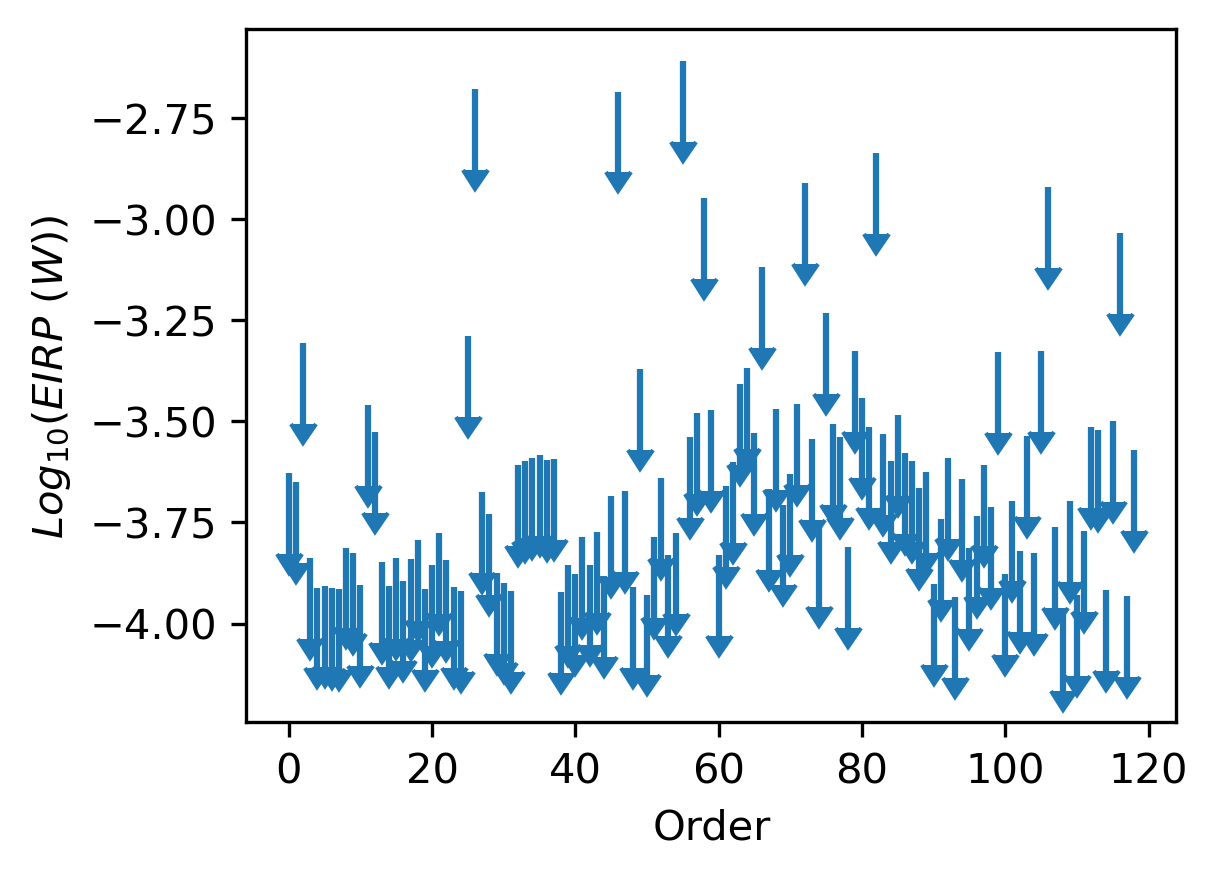}
\includegraphics[width=0.45\linewidth]{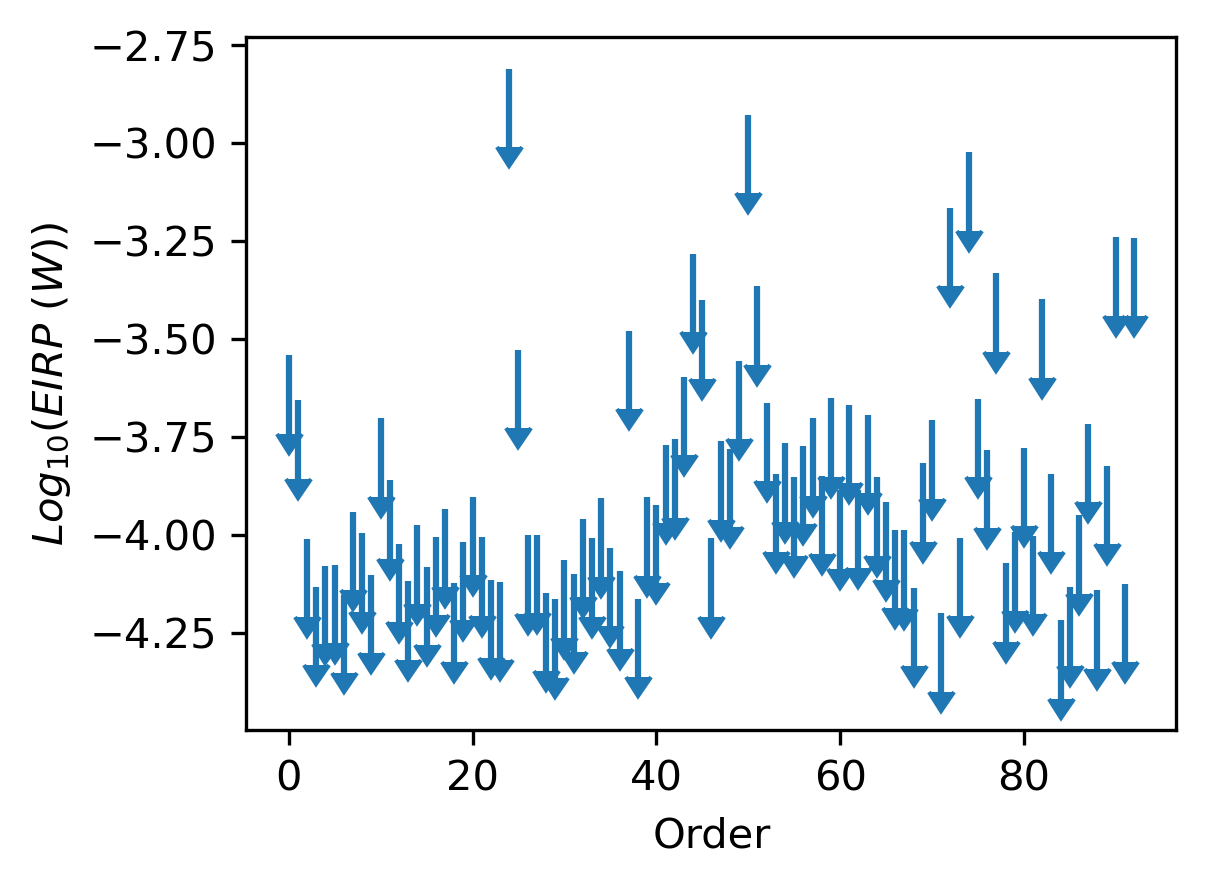}
\includegraphics[width=0.45\linewidth]{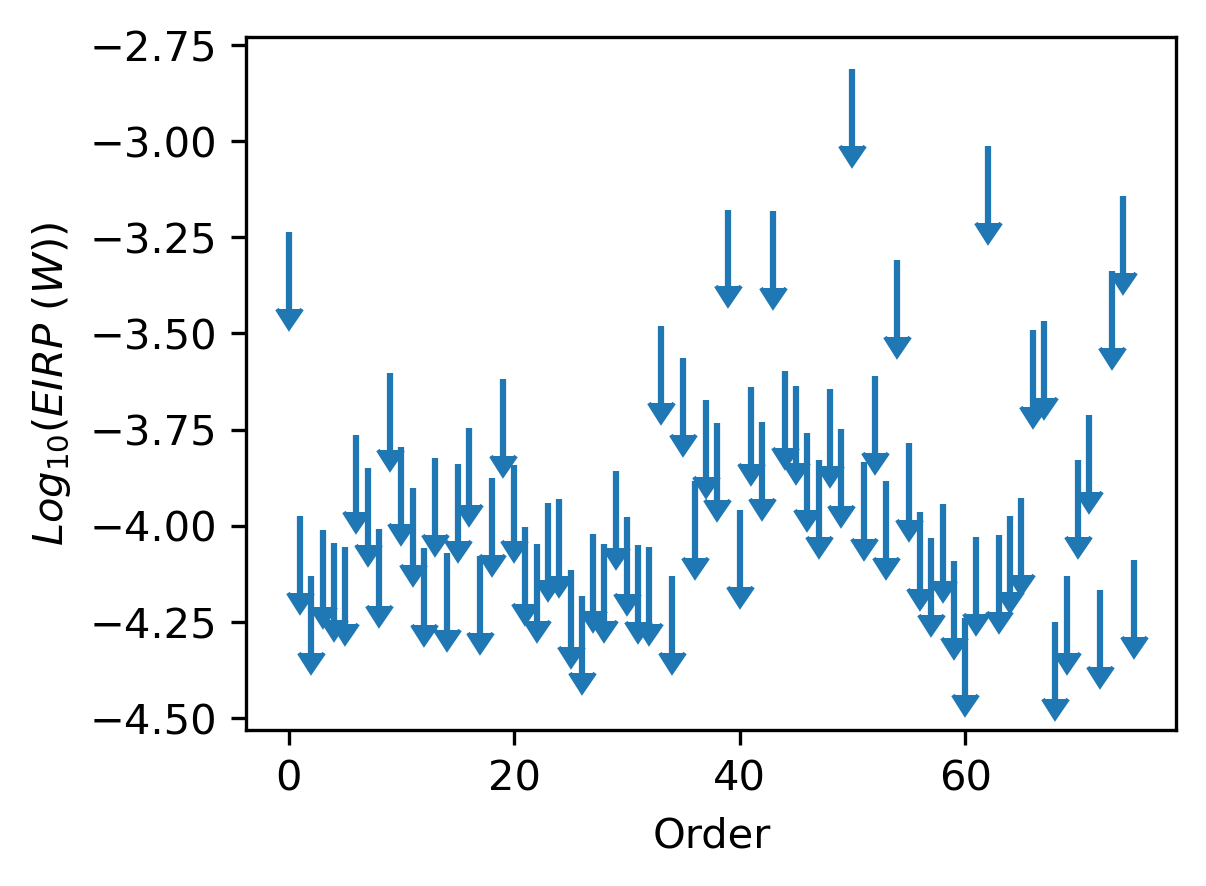}
\includegraphics[width=0.45\linewidth]{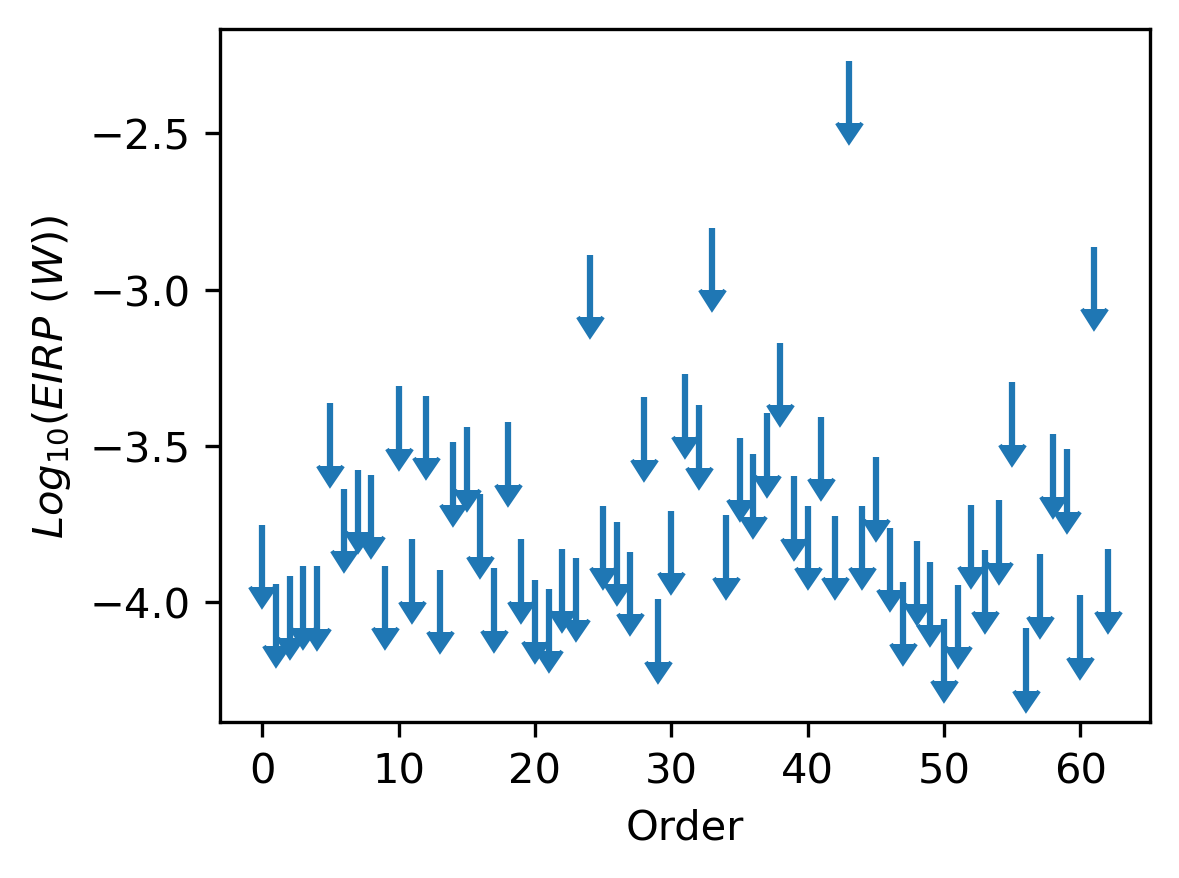}
\caption{EIRP upper limits ($4\sigma$) at 88 MHz to 216 MHz, from top left to bottom right.  The horizontal axis is labelled ``Order'', which is simply the order in which TLEs for satellites were returned from the space-track.org query.}
\label{fig8}
\end{figure*}

We calculate upper limits on the Equivalent Isotropic Radiated Power (EIRP) based on our observed intensity upper limits, for all objects at all frequencies, and show them in Figure \ref{fig8}.  From these results we see that, in general, we reach 4$\sigma$ upper limits on EIRP of well below 1 mW, and below 10 $\mu$W in the best cases (see details for each satellite in the Appendix).

We conclude that RFI due to unintended emissions from satellites in geostationary and geosynchronous orbits is currently likely to be undetectable at typical MWA sensitivities, based on the assumptions that we have made (emission of constant strength filling both the observing bandwidth and observing time).  There is currently little indication that this type of RFI will be an issue in the same way that unintended emissions from objects in LEO is becoming an issue.  

For comparison to previous work, detections of unintended emissions from objects in LEO by \cite{2023A&A...678L...6G} at levels of $\sim$200 Jy, in an approximately 1 MHz bandwidth, correspond to EIRPs of a few $\mu$W.  In the most sensitive of our data, we are approaching 4$\sigma$ EIRP limits at comparable levels for objects in GEO, a factor of $\sim$70 further away.  In terms of comparison of our data to industry Electromagnetic Compatibility (EMC) standards, \cite{2023A&A...676A..75D} notes that ``currently no international agency or space law ... requires a spacecraft to comply to a certain EMC standard''.  A range of standards can be used for comparison to our data, to give a feel for the data in an EMC sense.  We choose CISPR 32 Class A, for which the emission limit is 40 dB(uV/m) at a distance of 10 m, across the frequency range of 30 - 230 MHz, and measured in a 120 kHz band.  From this limit, we calculate the EIRP we would expect for the bandwidth used for our observations (assuming that the power density in frequency is the same for the 120 kHz band and our full observing bandwidth).  We calculate an equivalent CISPR 32 Class A EIRP of $1.5\times10^{-2}$ W, for our bandwidth, which is well above all the 4$\sigma$ upper limits presented in Figure \ref{fig8}.  The measurements we make indicate that any UEMR from GEO satellites we have observed is well below CISPR 32 Class A limits.

This is a welcome overall conclusion and the current study provides a useful baseline for comparison to future detailed monitoring of the RFI environment of the site of the SKA-Low, particularly as the SKA-Low sensitivity reaches and then significantly exceeds the current MWA sensitivity.  An additional factor of 50 sensitivity in the final SKA-Low would bring EIRP sensitivity to GEO objects into at least the $\sim$100nW range.  Given the emerging issues with objects in LEO, the situation for objects in GEO and other orbital domains should be monitored regularly into the future.

While we assume continuous, broadband signals for the purposes of deriving our limits, limits are also possible in the case that any signals are narrow band in nature.  In this case, a narrow band signal would be diluted by the full observed bandwidth such that our upper limits would be higher by the factor $\frac{32}{\Delta \nu}$, where $\Delta \nu$ is the width of the narrow band signal in MHz.  The GLEAM-X visibility data would be amenable to a possible future search for narrow band RFI from GEO satellites, as the inherent frequency resolution of the data is 10 kHz.  An analysis that leads to 10 kHz images, rather than 30.72 MHz images, is possible, but would require a factor of approximately 2000 more compute resource.

\begin{acknowledgement}
We thank two anonymous referees for their highly useful comments on our manuscript, which helped improve its quality, in particular their insights into spacecraft operations and EMC standards.  This scientific work uses data obtained from Inyarrimanha Ilgari Bundara, the CSIRO Murchison Radio-astronomy Observatory. We acknowledge the Wajarri Yamaji People as the Traditional Owners and Native Title Holders of the observatory site. Support for the operation of the MWA is provided by the Australian Government (NCRIS), under a contract to Curtin University administered by Astronomy Australia Limited. We acknowledge the Pawsey Supercomputing Centre which is supported by the Western Australian and Australian Governments.  N.H.-W. is the recipient of an Australian Research Council Future Fellowship (project number FT190100231).
\end{acknowledgement}


\printbibliography

@ARTICLE{2023A&A...678L...6G,
       author = {{Grigg}, D. and {Tingay}, S.~J. and {Sokolowski}, M. and {Wayth}, R.~B. and {Indermuehle}, B. and {Prabu}, S.},
        title = "{Detection of intended and unintended emissions from Starlink satellites in the SKA-Low frequency range, at the SKA-Low site, with an SKA-Low station analogue}",
      journal = {Astronomy and Astrophysics},
         year = 2023,
        month = oct,
       volume = {678},
          eid = {L6},
        pages = {L6},
          doi = {10.1051/0004-6361/202347654},
archivePrefix = {arXiv},
       eprint = {2309.15672},
 primaryClass = {astro-ph.IM},
       adsurl = {https://ui.adsabs.harvard.edu/abs/2023A&A...678L...6G}
}

@ARTICLE{2023A&A...676A..75D,
       author = {{Di Vruno}, F. and {Winkel}, B. and {Bassa}, C.~G. and {J{\'o}zsa}, G.~I.~G. and {Brentjens}, M.~A. and {Jessner}, A. and {Garrington}, S.},
        title = "{Unintended electromagnetic radiation from Starlink satellites detected with LOFAR between 110 and 188 MHz}",
      journal = {Astronomy and Astrophysics},
         year = 2023,
        month = aug,
       volume = {676},
          eid = {A75},
        pages = {A75},
          doi = {10.1051/0004-6361/202346374},
archivePrefix = {arXiv},
       eprint = {2307.02316},
 primaryClass = {astro-ph.IM},
       adsurl = {https://ui.adsabs.harvard.edu/abs/2023A&A...676A..75D}
}

@ARTICLE{2024arXiv241214483G,
       author = {{Grigg}, D. and {Tingay}, S.J. and {Prabu}, S. and {Sokolowski}, M. and {Indermuehle}, B.},
        title = "{Enhanced detection and identification of satellites using an all-sky multi-frequency survey with prototype SKA-Low stations}",
      journal = {arXiv e-prints},
         year = 2024,
        month = dec,
          eid = {arXiv:2412.14483},
        pages = {arXiv:2412.14483},
          doi = {10.48550/arXiv.2412.14483},
archivePrefix = {arXiv},
       eprint = {2412.14483},
 primaryClass = {astro-ph.EP},
       adsurl = {https://ui.adsabs.harvard.edu/abs/2024arXiv241214483G}
}

@ARTICLE{2024A&A...689L..10B,
       author = {{Bassa}, C.~G. and {Di Vruno}, F. and {Winkel}, B. and {J{\'o}zsa}, G.~I.~G. and {Brentjens}, M.~A. and {Zhang}, X.},
        title = "{Bright unintended electromagnetic radiation from second-generation Starlink satellites}",
      journal = {Astronomy and Astrophysics},
         year = 2024,
        month = sep,
       volume = {689},
          eid = {L10},
        pages = {L10},
          doi = {10.1051/0004-6361/202451856},
archivePrefix = {arXiv},
       eprint = {2409.11767},
 primaryClass = {astro-ph.IM},
       adsurl = {https://ui.adsabs.harvard.edu/abs/2024A&A...689L..10B}
}

@ARTICLE{2024AcAau.225.1019M,
       author = {{Marzioli}, Paolo and {Oltrogge}, Daniel and {Taiatu}, Claudiu Mihai and {Skinner}, Mark A. and {Court}, Andy},
        title = "{Management of radio-frequency interferences for space traffic management: Current regulations, operations practice, technology mitigation solutions and future trends}",
      journal = {Acta Astronautica},
         year = 2024,
        month = dec,
       volume = {225},
        pages = {1019-1030},
          doi = {10.1016/j.actaastro.2024.10.005},
       adsurl = {https://ui.adsabs.harvard.edu/abs/2024AcAau.225.1019M}
}

@ARTICLE{2023PASA...40...56P,
       author = {{Prabu}, S. and {Tingay}, S.~J. and {Williams}, A.},
        title = "{A near-field treatment of aperture synthesis techniques using the Murchison Widefield Array}",
      journal = {Publications of the Astronomical Society of Australia},
         year = 2023,
        month = dec,
       volume = {40},
          eid = {e056},
        pages = {e056},
          doi = {10.1017/pasa.2023.56},
archivePrefix = {arXiv},
       eprint = {2310.19320},
 primaryClass = {astro-ph.IM},
       adsurl = {https://ui.adsabs.harvard.edu/abs/2023PASA...40...56P}
}

@ARTICLE{2022AdSpR..70..812P,
       author = {{Prabu}, S. and {Hancock}, P. and {Zhang}, X. and {Tingay}, S.~J. and {Hodgson}, T. and {Crosse}, B. and {Johnston-Hollitt}, M.},
        title = "{Improved sensitivity for space domain awareness observations with the murchison widefield array}",
      journal = {Advances in Space Research},
         year = 2022,
        month = aug,
       volume = {70},
       number = {3},
        pages = {812-824},
          doi = {10.1016/j.asr.2022.05.013},
archivePrefix = {arXiv},
       eprint = {2205.05868},
 primaryClass = {astro-ph.IM},
       adsurl = {https://ui.adsabs.harvard.edu/abs/2022AdSpR..70..812P}
}

@ARTICLE{2020MNRAS.498..265W,
       author = {{Wilensky}, Michael J. and {Barry}, Nichole and {Morales}, Miguel F. and {Hazelton}, Bryna J. and {Byrne}, Ruby},
        title = "{Quantifying excess power from radio frequency interference in Epoch of Reionization measurements}",
      journal = {Monthly Notices of the Royal Astronomical Society},
         year = 2020,
        month = oct,
       volume = {498},
       number = {1},
        pages = {265-275},
          doi = {10.1093/mnras/staa2442},
archivePrefix = {arXiv},
       eprint = {2004.07819},
 primaryClass = {astro-ph.IM},
       adsurl = {https://ui.adsabs.harvard.edu/abs/2020MNRAS.498..265W}
}

@ARTICLE{2023MNRAS.524.3231F,
       author = {{Finlay}, Chris and {Bassett}, Bruce A. and {Kunz}, Martin and {Oozeer}, Nadeem},
        title = "{Trajectory-based RFI subtraction and calibration for radio interferometry}",
      journal = {Monthly Notices of the Royal Astronomical Society},
         year = 2023,
        month = sep,
       volume = {524},
       number = {3},
        pages = {3231-3251},
          doi = {10.1093/mnras/stad1979},
archivePrefix = {arXiv},
       eprint = {2301.04188},
 primaryClass = {astro-ph.IM},
       adsurl = {https://ui.adsabs.harvard.edu/abs/2023MNRAS.524.3231F}
}

@ARTICLE{5136190,

  author={Dewdney, Peter E. and Hall, Peter J. and Schilizzi, Richard T. and Lazio, T. Joseph L. W.},

  journal={Proceedings of the IEEE}, 

  title={The Square Kilometre Array}, 

  year={2009},

  volume={97},

  number={8},

  pages={1482-1496},

  doi={10.1109/JPROC.2009.2021005}}

@ARTICLE{2013PASA...30....7T,
       author = {{Tingay}, S.~J. and {Goeke}, R. and {Bowman}, J.~D. and {Emrich}, D. and {Ord}, S.~M. and {Mitchell}, D.~A. and {Morales}, M.~F. and {Booler}, T. and {Crosse}, B. and {Wayth}, R.~B. and {Lonsdale}, C.~J. and {Tremblay}, S. and {Pallot}, D. and {Colegate}, T. and {Wicenec}, A. and {Kudryavtseva}, N. and {Arcus}, W. and {Barnes}, D. and {Bernardi}, G. and {Briggs}, F. and {Burns}, S. and {Bunton}, J.~D. and {Cappallo}, R.~J. and {Corey}, B.~E. and {Deshpande}, A. and {Desouza}, L. and {Gaensler}, B.~M. and {Greenhill}, L.~J. and {Hall}, P.~J. and {Hazelton}, B.~J. and {Herne}, D. and {Hewitt}, J.~N. and {Johnston-Hollitt}, M. and {Kaplan}, D.~L. and {Kasper}, J.~C. and {Kincaid}, B.~B. and {Koenig}, R. and {Kratzenberg}, E. and {Lynch}, M.~J. and {Mckinley}, B. and {Mcwhirter}, S.~R. and {Morgan}, E. and {Oberoi}, D. and {Pathikulangara}, J. and {Prabu}, T. and {Remillard}, R.~A. and {Rogers}, A.~E.~E. and {Roshi}, A. and {Salah}, J.~E. and {Sault}, R.~J. and {Udaya-Shankar}, N. and {Schlagenhaufer}, F. and {Srivani}, K.~S. and {Stevens}, J. and {Subrahmanyan}, R. and {Waterson}, M. and {Webster}, R.~L. and {Whitney}, A.~R. and {Williams}, A. and {Williams}, C.~L. and {Wyithe}, J.~S.~B.},
        title = "{The Murchison Widefield Array: The Square Kilometre Array Precursor at Low Radio Frequencies}",
      journal = {Publications of the Astronomical Society of Australia},
         year = 2013,
        month = jan,
       volume = {30},
          eid = {e007},
        pages = {e007},
          doi = {10.1017/pasa.2012.007},
archivePrefix = {arXiv},
       eprint = {1206.6945},
 primaryClass = {astro-ph.IM},
       adsurl = {https://ui.adsabs.harvard.edu/abs/2013PASA...30....7T}
}

@ARTICLE{2018PASA...35...33W,
       author = {{Wayth}, Randall B. and {Tingay}, Steven J. and {Trott}, Cathryn M. and {Emrich}, David and {Johnston-Hollitt}, Melanie and {McKinley}, Ben and {Gaensler}, B.~M. and {Beardsley}, A.~P. and {Booler}, T. and {Crosse}, B. and {Franzen}, T.~M.~O. and {Horsley}, L. and {Kaplan}, D.~L. and {Kenney}, D. and {Morales}, M.~F. and {Pallot}, D. and {Sleap}, G. and {Steele}, K. and {Walker}, M. and {Williams}, A. and {Wu}, C. and {Cairns}, Iver. H. and {Filipovic}, M.~D. and {Johnston}, S. and {Murphy}, T. and {Quinn}, P. and {Staveley-Smith}, L. and {Webster}, R. and {Wyithe}, J.~S.~B.},
        title = "{The Phase II Murchison Widefield Array: Design overview}",
      journal = {Publications of the Astronomical Society of Australia},
         year = 2018,
        month = nov,
       volume = {35},
          eid = {e033},
        pages = {e033},
          doi = {10.1017/pasa.2018.37},
archivePrefix = {arXiv},
       eprint = {1809.06466},
 primaryClass = {astro-ph.IM},
       adsurl = {https://ui.adsabs.harvard.edu/abs/2018PASA...35...33W}
}

@ARTICLE{2024ApJ...976L..21H,
       author = {{Hurley-Walker}, N. and {McSweeney}, S.~J. and {Bahramian}, A. and {Rea}, N. and {Horv{\'a}th}, C. and {Buchner}, S. and {Williams}, A. and {Meyers}, B.~W. and {Strader}, Jay and {Aydi}, Elias and {Urquhart}, Ryan and {Chomiuk}, Laura and {Galvin}, T.~J. and {Coti Zelati}, F. and {Bailes}, Matthew},
        title = "{A 2.9 hr Periodic Radio Transient with an Optical Counterpart}",
      journal = {The Astrophysical Journal Letters},
         year = 2024,
        month = dec,
       volume = {976},
       number = {2},
          eid = {L21},
        pages = {L21},
          doi = {10.3847/2041-8213/ad890e},
archivePrefix = {arXiv},
       eprint = {2408.15757},
 primaryClass = {astro-ph.SR},
       adsurl = {https://ui.adsabs.harvard.edu/abs/2024ApJ...976L..21H}
}

@INPROCEEDINGS{Racelis2018-pj,
  title      = "Correction: High-integrity {TLE} error models for {MEO} and
                {GEO} satellites",
  booktitle  = "2018 {AIAA} {SPACE} and Astronautics Forum and Exposition",
  author     = "Racelis, Danielle and Joerger, Mathieu",
  publisher  = "American Institute of Aeronautics and Astronautics",
  month      =  sep,
  year       =  2018,
  address    = "Reston, Virginia",
  conference = "2018 AIAA SPACE and Astronautics Forum and Exposition",
  location   = "Orlando, FL"
}

@ARTICLE{2018PASA...35...11H,
       author = {{Hancock}, Paul J. and {Trott}, Cathryn M. and {Hurley-Walker}, Natasha},
        title = "{Source Finding in the Era of the SKA (Precursors): Aegean 2.0}",
      journal = {Publications of the Astronomical Society of Australia},
         year = 2018,
        month = mar,
       volume = {35},
          eid = {e011},
        pages = {e011},
          doi = {10.1017/pasa.2018.3},
archivePrefix = {arXiv},
       eprint = {1801.05548},
 primaryClass = {astro-ph.IM},
       adsurl = {https://ui.adsabs.harvard.edu/abs/2018PASA...35...11H}
}

@ARTICLE{2024PASA...41...54R,
       author = {{Ross}, Kathryn and {Hurley-Walker}, Natasha and {Galvin}, Timothy James and {Venville}, Brandon and {Duchesne}, Stefan William and {Morgan}, John and {An}, Tao and {G{\"u}rkan}, Gulay and {Hancock}, Paul J. and {Heald}, George and {Johnston-Hollitt}, Melanie and {White}, Sarah V.},
        title = "{GaLactic and Extragalactic All-sky Murchison Widefield Array eXtended (GLEAM-X) survey II: Second Data Release}",
      journal = {Publications of the Astronomical Society of Australia},
         year = 2024,
        month = sep,
       volume = {41},
          eid = {e054},
        pages = {e054},
          doi = {10.1017/pasa.2024.57},
archivePrefix = {arXiv},
       eprint = {2406.06921},
 primaryClass = {astro-ph.GA},
       adsurl = {https://ui.adsabs.harvard.edu/abs/2024PASA...41...54R}
}

@ARTICLE{2018MNRAS.473.2965M,
       author = {{Morgan}, J.~S. and {Macquart}, J. -P. and {Ekers}, R. and {Chhetri}, R. and {Tokumaru}, M. and {Manoharan}, P.~K. and {Tremblay}, S. and {Bisi}, M.~M. and {Jackson}, B.~V.},
        title = "{Interplanetary Scintillation with the Murchison Widefield Array I: a sub-arcsecond survey over 900 deg$^{2}$ at 79 and 158 MHz}",
      journal = {Monthly Notices of the Royal Astronomical Society},
         year = 2018,
        month = jan,
       volume = {473},
       number = {3},
        pages = {2965-2983},
          doi = {10.1093/mnras/stx2284},
archivePrefix = {arXiv},
       eprint = {1709.00312},
 primaryClass = {astro-ph.IM},
       adsurl = {https://ui.adsabs.harvard.edu/abs/2018MNRAS.473.2965M}
}

@ARTICLE{2025PASA...42...10D,
       author = {{Ducharme}, Jade M. and {Pober}, Jonathan C.},
        title = "{Altitude estimation of radio frequency interference sources via interferometric near-field corrections}",
      journal = {Publications of the Astronomical Society of Australia},
         year = 2025,
        month = feb,
       volume = {42},
          eid = {e010},
        pages = {e010},
          doi = {10.1017/pasa.2024.123},
archivePrefix = {arXiv},
       eprint = {2502.08867},
 primaryClass = {astro-ph.IM},
       adsurl = {https://ui.adsabs.harvard.edu/abs/2025PASA...42...10D}
}

@ARTICLE{2025A&A...699A.307G,
       author = {{Grigg}, D. and {Tingay}, S.~J. and {Sokolowski}, M.},
        title = "{The growing impact of unintended Starlink broadband emission on radio astronomy in the SKA-Low frequency range}",
      journal = {Astronomy and Astrophysics},
         year = 2025,
        month = jul,
       volume = {699},
          eid = {A307},
        pages = {A307},
          doi = {10.1051/0004-6361/202554787},
archivePrefix = {arXiv},
       eprint = {2506.02831},
 primaryClass = {astro-ph.IM},
       adsurl = {https://ui.adsabs.harvard.edu/abs/2025A&A...699A.307G}
}

@ARTICLE{2014MNRAS.444..606O,
       author = {{Offringa}, A.~R. and {McKinley}, B. and {Hurley-Walker}, N. and {Briggs}, F.~H. and {Wayth}, R.~B. and {Kaplan}, D.~L. and {Bell}, M.~E. and {Feng}, L. and {Neben}, A.~R. and {Hughes}, J.~D. and {Rhee}, J. and {Murphy}, T. and {Bhat}, N.~D.~R. and {Bernardi}, G. and {Bowman}, J.~D. and {Cappallo}, R.~J. and {Corey}, B.~E. and {Deshpande}, A.~A. and {Emrich}, D. and {Ewall-Wice}, A. and {Gaensler}, B.~M. and {Goeke}, R. and {Greenhill}, L.~J. and {Hazelton}, B.~J. and {Hindson}, L. and {Johnston-Hollitt}, M. and {Jacobs}, D.~C. and {Kasper}, J.~C. and {Kratzenberg}, E. and {Lenc}, E. and {Lonsdale}, C.~J. and {Lynch}, M.~J. and {McWhirter}, S.~R. and {Mitchell}, D.~A. and {Morales}, M.~F. and {Morgan}, E. and {Kudryavtseva}, N. and {Oberoi}, D. and {Ord}, S.~M. and {Pindor}, B. and {Procopio}, P. and {Prabu}, T. and {Riding}, J. and {Roshi}, D.~A. and {Shankar}, N. Udaya and {Srivani}, K.~S. and {Subrahmanyan}, R. and {Tingay}, S.~J. and {Waterson}, M. and {Webster}, R.~L. and {Whitney}, A.~R. and {Williams}, A. and {Williams}, C.~L.},
        title = "{WSCLEAN: an implementation of a fast, generic wide-field imager for radio astronomy}",
      journal = {Monthly Notices of the Royal Astronomical Society},
         year = 2014,
        month = oct,
       volume = {444},
       number = {1},
        pages = {606-619},
          doi = {10.1093/mnras/stu1368},
archivePrefix = {arXiv},
       eprint = {1407.1943},
 primaryClass = {astro-ph.IM},
       adsurl = {https://ui.adsabs.harvard.edu/abs/2014MNRAS.444..606O}
}

@ARTICLE{2019ascl.soft07024R,
       author = {{Rhodes}, Brandon},
        title = "{Skyfield: High precision research-grade positions for planets and Earth satellites generator}",
 howpublished = {Astrophysics Source Code Library, record ascl:1907.024},
         year = 2019,
        month = jul,
          eid = {ascl:1907.024},
archivePrefix = {ascl},
       eprint = {1907.024},
       adsurl = {https://ui.adsabs.harvard.edu/abs/2019ascl.soft07024R}
}

@ARTICLE{2025A&A...698A.244Z,
       author = {{Zhang}, X. and {Zarka}, P. and {Viou}, C. and {Loh}, A. and {Bassa}, C.~G. and {Duchene}, Q. and {Tasse}, C. and {Grie{\ss}meier}, J.-M. and {Turner}, J.~D. and {Ulyanov}, O. and {Koopmans}, L.~V.~E. and {Mertens}, F. and {Zakharenko}, V. and {Briand}, C. and {Cecconi}, B. and {Vermeulen}, R. and {Konovalenko}, O. and {Girard}, J.~N. and {Corbel}, S.},
        title = "{Broadband polarized radio emission detected from Starlink satellites below 100 MHz with NenuFAR}",
      journal = {Astronomy and Astrophysics},
         year = 2025,
        month = jun,
       volume = {698},
          eid = {A244},
        pages = {A244},
          doi = {10.1051/0004-6361/202554152},
archivePrefix = {arXiv},
       eprint = {2504.10032},
 primaryClass = {astro-ph.IM},
       adsurl = {https://ui.adsabs.harvard.edu/abs/2025A&A...698A.244Z}
}

@ARTICLE{2025PASA...42..129H,
       author = {{Horv{\'a}th}, Csan{\'a}d and {Hurley-Walker}, Natasha and {McSweeney}, Samuel and {Galvin}, Timothy James and {Morgan}, John},
        title = "{A long period transient search method for the Murchison Widefield Array}",
      journal = {Publications of the Astronomical Society of Australia},
         year = 2025,
        month = sep,
       volume = {42},
          eid = {e129},
        pages = {e129},
          doi = {10.1017/pasa.2025.10093},
archivePrefix = {arXiv},
       eprint = {2509.06315},
 primaryClass = {astro-ph.IM},
       adsurl = {https://ui.adsabs.harvard.edu/abs/2025PASA...42..129H}
}

@ARTICLE{2025ApJ...981..143M,
       author = {{Mcsweeney}, Samuel J. and {Moseley}, Jared and {Hurley-Walker}, Natasha and {Grover}, Garvit and {Horv{\'a}th}, Csan{\'a}d and {Galvin}, Timothy J. and {Meyers}, Bradley W. and {Tan}, Chia Min},
        title = "{Discovery of an RRAT-like Pulsar via Its Single Pulses in a Murchison Widefield Array Imaging Survey}",
      journal = {The Astrophysical Journal},
         year = 2025,
        month = mar,
       volume = {981},
       number = {2},
          eid = {143},
        pages = {143},
          doi = {10.3847/1538-4357/adb27f},
archivePrefix = {arXiv},
       eprint = {2502.02130},
 primaryClass = {astro-ph.HE},
       adsurl = {https://ui.adsabs.harvard.edu/abs/2025ApJ...981..143M}
}

\appendix

\section{Upper limits for unintended emission at 88, 118, 154, 185, and 216 MHz}

\begin{center}
\begin{longtable}{|c|c|l|l|l|l|l|}
\caption{Upper limits at 88 MHz} \label{tab2} \\

\hline \multicolumn{1}{|c|}{\textbf{NORAD ID \#}} & \multicolumn{1}{c|}{\textbf{Integration time (s)}} & \multicolumn{1}{c|}{\textbf{Pixel intensity (Jy/beam)}} & \multicolumn{1}{c|}{\textbf{RMS (Jy/beam)}} & \multicolumn{1}{c|}{\textbf{Mean distance (km)}} & \multicolumn{1}{c|}{\textbf{$Log_{10} (EIRP (W))$}}\\ \hline 
\endfirsthead

\multicolumn{3}{c}%
{{\bfseries \tablename\ \thetable{} -- continued from previous page}} \\
\hline \multicolumn{1}{|c|}{\textbf{NORAD ID \#}} & \multicolumn{1}{c|}{\textbf{Integration time (s)}} & \multicolumn{1}{c|}{\textbf{Pixel intensity (Jy/beam)}} & \multicolumn{1}{c|}{\textbf{RMS (Jy/beam)}} & \multicolumn{1}{c|}{\textbf{Mean distance (km)}} & \multicolumn{1}{c|}{\textbf{$Log_{10} (EIRP (W))$}}\\ \hline 
\endhead

\hline \multicolumn{3}{|r|}{{Continued on next page}} \\ \hline
\endfoot

\hline \hline
\endlastfoot
2217&4976&0.04&0.029&34806&-2.9 \\ 
5854&5000&0.0729&0.026&36514&-2.9 \\ 
7324&5000&0.1104&0.051&37094&-2.6 \\ 
8357&5000&0.1029&0.03&36916&-2.8 \\ 
8476&5000&0.0749&0.031&37340&-2.8 \\ 
10516&5000&0.0623&0.03&37465&-2.8 \\ 
10792&5000&0.0859&0.051&37246&-2.6 \\ 
11484&5000&0.0717&0.028&37111&-2.8 \\ 
11567&5000&0.0749&0.034&36651&-2.8 \\ 
11570&5000&0.0756&0.034&37768&-2.7 \\ 
11571&5000&0.0937&0.034&36714&-2.8 \\ 
11941&5000&0.1025&0.049&37982&-2.6 \\ 
12618&5000&0.0711&0.04&37134&-2.7 \\ 
12635&5000&0.0819&0.037&37796&-2.7 \\ 
13086&5000&0.0472&0.029&37685&-2.8 \\ 
13089&2080&0.139&0.051&37292&-2.6 \\ 
13651&5000&0.0896&0.051&37375&-2.6 \\ 
13878&1060&0.1989&0.094&37922&-2.3 \\ 
14193&5000&0.1593&0.058&36954&-2.5 \\ 
14786&5000&0.0621&0.035&38081&-2.7 \\ 
14948&5000&0.0646&0.035&37252&-2.7 \\ 
15236&5000&0.0571&0.036&38146&-2.7 \\ 
15545&5000&0.0651&0.031&37056&-2.8 \\ 
15581&5000&0.1406&0.047&37598&-2.6 \\ 
15826&5000&0.0321&0.022&37007&-2.9 \\ 
16339&120&0.5182&0.158&37205&-2.1 \\ 
16649&4096&0.118&0.071&37575&-2.4 \\ 
17125&5000&0.0703&0.04&37215&-2.7 \\ 
18316&5000&0.0679&0.04&37272&-2.7 \\ 
18570&5000&0.0481&0.024&37235&-2.9 \\ 
19090&5000&0.0987&0.032&38472&-2.7 \\ 
19094&1568&0.5386&0.215&37278&-1.9 \\ 
19710&5000&0.0301&0.02&36791&-3.0 \\ 
19928&5000&0.0411&0.025&36950&-2.9 \\ 
20263&5000&0.0536&0.03&36942&-2.8 \\ 
20696&5000&0.0496&0.035&37036&-2.7 \\ 
20705&5000&0.0341&0.019&37219&-3.0 \\ 
20872&5000&0.0626&0.033&37137&-2.8 \\ 
21132&5000&0.0473&0.023&36856&-2.9 \\ 
21139&5000&0.0798&0.026&37039&-2.9 \\ 
21533&2484&0.2023&0.056&36814&-2.5 \\ 
21762&5000&0.0534&0.026&37153&-2.9 \\ 
21925&5000&0.0493&0.026&36678&-2.9 \\ 
22115&5000&0.0608&0.04&37334&-2.7 \\ 
22210&5000&0.0919&0.038&36939&-2.7 \\ 
22557&5000&0.05&0.04&37164&-2.7 \\ 
22883&5000&0.0269&0.02&36732&-3.0 \\ 
22916&5000&0.0546&0.037&36903&-2.7 \\ 
23111&5000&0.0544&0.022&35908&-3.0 \\ 
23171&5000&0.0717&0.033&37065&-2.8 \\ 
23571&5000&0.062&0.034&36948&-2.8 \\ 
23628&5000&0.0432&0.023&36585&-2.9 \\ 
23649&2196&0.0882&0.045&37647&-2.6 \\ 
23651&5000&0.0432&0.016&36742&-3.1 \\ 
23720&5000&0.0594&0.026&37002&-2.9 \\ 
23775&5000&0.0773&0.021&36868&-3.0 \\ 
23880&5000&0.0567&0.022&36717&-2.9 \\ 
23949&5000&0.0267&0.017&36448&-3.1 \\ 
24653&5000&0.0635&0.042&37023&-2.7 \\ 
24798&5000&0.0698&0.022&36708&-3.0 \\ 
25311&2472&0.0857&0.044&37775&-2.6 \\ 
25501&4072&0.0619&0.039&37143&-2.7 \\ 
25558&5000&0.0561&0.02&36900&-3.0 \\ 
25894&5000&0.039&0.021&36593&-3.0 \\ 
25937&5000&0.0539&0.024&36870&-2.9 \\ 
26107&5000&0.0406&0.026&36555&-2.9 \\ 
26108&5000&0.0925&0.036&36800&-2.7 \\ 
26480&5000&0.0297&0.02&37112&-3.0 \\ 
26559&5000&0.0451&0.019&36834&-3.0 \\ 
26575&1620&0.0392&0.042&37202&-2.7 \\ 
26892&5000&0.0371&0.016&36547&-3.1 \\ 
26895&5000&0.0324&0.017&36492&-3.1 \\ 
26985&5000&0.0532&0.022&36872&-2.9 \\ 
27169&5000&0.0347&0.019&36604&-3.0 \\ 
27399&5000&0.0541&0.019&36438&-3.0 \\ 
27683&5000&0.0873&0.025&36954&-2.9 \\ 
27712&5000&0.0576&0.023&36600&-2.9 \\ 
27780&5000&0.0423&0.019&36496&-3.0 \\ 
27820&5000&0.0707&0.026&36889&-2.9 \\ 
28139&924&0.1598&0.106&36466&-2.3 \\ 
28161&5000&0.0585&0.024&36761&-2.9 \\ 
28707&5000&0.0646&0.031&36669&-2.8 \\ 
28786&5000&0.0617&0.022&36605&-2.9 \\ 
29045&5000&0.0979&0.044&36844&-2.6 \\ 
29272&5000&0.0957&0.033&36735&-2.8 \\ 
29349&5000&0.0478&0.022&36596&-2.9 \\ 
29398&5000&0.0601&0.025&36646&-2.9 \\ 
29640&5000&0.0785&0.035&37275&-2.7 \\ 
31800&5000&0.0392&0.022&36596&-3.0 \\ 
32019&5000&0.0377&0.024&36631&-2.9 \\ 
32404&2256&0.1642&0.06&36857&-2.5 \\ 
32767&5000&0.0709&0.044&36843&-2.6 \\ 
33510&5000&0.0708&0.048&36506&-2.6 \\ 
34941&5000&0.0492&0.026&36642&-2.9 \\ 
35696&5000&0.0693&0.042&36827&-2.7 \\ 
35812&5000&0.0738&0.022&36581&-2.9 \\ 
36744&5000&0.0636&0.033&36734&-2.8 \\ 
37150&5000&0.0652&0.025&36672&-2.9 \\ 
37207&5000&0.0295&0.024&36623&-2.9 \\ 
37234&5000&0.059&0.026&36657&-2.9 \\ 
37256&700&0.2227&0.107&37805&-2.2 \\ 
37265&5000&0.0495&0.022&36587&-3.0 \\ 
37384&1700&0.0641&0.027&37190&-2.8 \\ 
37677&5000&0.0455&0.023&36632&-2.9 \\ 
37763&764&0.6649&0.223&37676&-1.9 \\ 
37776&5000&0.0528&0.024&36618&-2.9 \\ 
37804&5000&0.0723&0.036&36778&-2.7 \\ 
37933&5000&0.1039&0.028&36695&-2.8 \\ 
37948&400&0.1906&0.071&36649&-2.4 \\ 
38049&5000&0.0512&0.02&36506&-3.0 \\ 
38331&5000&0.0401&0.025&36656&-2.9 \\ 
38332&5000&0.1029&0.045&36838&-2.6 \\ 
38694&5000&0.0705&0.035&36046&-2.8 \\ 
39256&5000&0.0554&0.022&36616&-2.9 \\ 
40141&5000&0.0345&0.022&36603&-3.0 \\ 
40367&5000&0.118&0.044&36761&-2.6 \\ 
40547&2400&0.0635&0.042&37641&-2.6 \\ 
40549&500&0.0855&0.049&36497&-2.6 \\ 
40872&5000&0.0506&0.02&35983&-3.0 \\ 
40896&5000&0.0627&0.024&37249&-2.9 \\ 
40990&5000&0.051&0.021&36088&-3.0 \\ 
41021&5000&0.0827&0.033&36740&-2.8 \\ 
41034&5000&0.0754&0.033&36742&-2.8 \\ 
41194&5000&0.0574&0.028&36694&-2.8 \\ 
41241&2100&0.0286&0.012&36057&-3.2 \\ 
41380&5000&0.0539&0.025&36646&-2.9 \\ 
41434&712&0.0995&0.027&36374&-2.9 \\ 
41469&5000&0.0586&0.034&36773&-2.8 \\ 
41586&5000&0.0417&0.021&36542&-3.0 \\ 
41725&5000&0.0974&0.04&36848&-2.7 \\ 
41745&5000&0.0623&0.026&36686&-2.9 \\ 
41838&5000&0.0346&0.021&36555&-3.0 \\ 
41882&5000&0.0535&0.03&36715&-2.8 \\ 
41903&5000&0.0639&0.024&36623&-2.9 \\ 
41911&5000&0.0537&0.026&36630&-2.9 \\ 
41944&5000&0.0367&0.022&36595&-3.0 \\ 
42662&5000&0.0396&0.022&36629&-2.9 \\ 
42691&5000&0.0319&0.022&36590&-2.9 \\ 
42763&5000&0.1067&0.039&36802&-2.7 \\ 
42917&5000&0.0578&0.029&36711&-2.8 \\ 
42942&5000&0.0477&0.023&36629&-2.9 \\ 
42951&5000&0.0634&0.024&36618&-2.9 \\ 
42984&5000&0.0362&0.022&36596&-2.9 \\ 
43339&2884&0.1262&0.069&36909&-2.4 \\ 
43432&5000&0.1368&0.033&36726&-2.8 \\ 
43450&3644&0.0962&0.062&36914&-2.5 \\ 
43463&5000&0.0464&0.022&36601&-3.0 \\ 
43539&1900&0.2508&0.093&36812&-2.3 \\ 
43587&5000&0.0742&0.026&36651&-2.9 \\ 
43823&5000&0.0806&0.033&36735&-2.8 \\ 
43874&5000&0.0526&0.023&36615&-2.9 \\ 
43920&5000&0.0916&0.037&36779&-2.7 \\ 
44067&5000&0.0681&0.036&36791&-2.7 \\ 
44204&1900&0.0891&0.03&36904&-2.8 \\ 
44337&1600&0.1087&0.043&36991&-2.7 \\ 
44709&900&0.0944&0.071&36988&-2.4 \\ 
44910&5000&0.0797&0.026&36708&-2.9 \\ 
45246&5000&0.0944&0.033&36741&-2.8 \\ 
45807&5000&0.0688&0.023&36728&-2.9 \\ 
45863&4044&0.1087&0.061&36907&-2.5 \\ 
45920&5000&0.0511&0.021&36588&-3.0 \\ 
46112&5000&0.0548&0.024&36624&-2.9 \\

\end{longtable}
\end{center}

\begin{center}
\begin{longtable}{|c|c|l|l|l|l|l|}
\caption{Upper limits at 118 MHz}  \label{tab3} \\

\hline \multicolumn{1}{|c|}{\textbf{NORAD ID \#}} & \multicolumn{1}{c|}{\textbf{Integration time (s)}} & \multicolumn{1}{c|}{\textbf{Pixel intensity (Jy/beam)}} & \multicolumn{1}{c|}{\textbf{RMS (Jy/beam)}} & \multicolumn{1}{c|}{\textbf{Mean distance (km)}} & \multicolumn{1}{c|}{\textbf{$Log_{10} (EIRP (W))$}}\\ \hline 
\endfirsthead

\multicolumn{3}{c}%
{{\bfseries \tablename\ \thetable{} -- continued from previous page}} \\
\hline \multicolumn{1}{|c|}{\textbf{NORAD ID \#}} & \multicolumn{1}{c|}{\textbf{Integration time (s)}} & \multicolumn{1}{c|}{\textbf{Pixel intensity (Jy/beam)}} & \multicolumn{1}{c|}{\textbf{RMS (Jy/beam)}} & \multicolumn{1}{c|}{\textbf{Mean distance (km)}} & \multicolumn{1}{c|}{\textbf{$Log_{10} (EIRP (W))$}}\\ \hline 
\endhead

\hline \multicolumn{3}{|r|}{{Continued on next page}} \\ \hline
\endfoot

\hline \hline
\endlastfoot
2217&2572&0.0189&0.007&34611&-3.5 \\ 
5854&5100&0.0086&0.004&36513&-3.7 \\ 
7324&1632&0.0441&0.018&37109&-3.0 \\ 
8357&5100&0.0092&0.004&36925&-3.7 \\ 
8476&5100&0.0133&0.006&37363&-3.5 \\ 
10516&5100&0.017&0.006&37487&-3.5 \\ 
11484&5100&0.0214&0.009&37135&-3.3 \\ 
11567&5100&0.0117&0.006&36662&-3.6 \\ 
11570&5100&0.0126&0.005&37784&-3.6 \\ 
11571&5100&0.0097&0.005&36724&-3.6 \\ 
11941&524&0.0695&0.029&37422&-2.8 \\ 
12618&4340&0.0205&0.012&37167&-3.2 \\ 
12635&2700&0.014&0.007&37454&-3.4 \\ 
13086&5100&0.0112&0.007&37710&-3.4 \\ 
14786&2992&0.0258&0.01&38042&-3.3 \\ 
14948&5100&0.0138&0.008&37274&-3.4 \\ 
15236&4180&0.0406&0.009&38090&-3.3 \\ 
15545&3400&0.0185&0.007&36769&-3.4 \\ 
15581&4424&0.0312&0.014&37705&-3.1 \\ 
15826&4200&0.0137&0.006&36875&-3.5 \\ 
17125&5100&0.0267&0.011&37238&-3.2 \\ 
18570&5100&0.0139&0.007&37252&-3.4 \\ 
19090&2580&0.0152&0.007&37958&-3.4 \\ 
19710&4696&0.0158&0.006&36742&-3.5 \\ 
19928&5100&0.0202&0.008&36972&-3.4 \\ 
20263&4996&0.0222&0.012&36960&-3.2 \\ 
20705&4900&0.0144&0.006&37202&-3.5 \\ 
21132&4400&0.0088&0.006&36758&-3.5 \\ 
21762&4100&0.0208&0.007&37021&-3.4 \\ 
21925&4364&0.0115&0.007&36584&-3.4 \\ 
22115&2292&0.0679&0.034&37789&-2.7 \\ 
22557&2392&0.0693&0.028&37535&-2.8 \\ 
22883&4500&0.0113&0.005&36650&-3.6 \\ 
23111&5100&0.0134&0.007&35925&-3.4 \\ 
23628&5100&0.0172&0.007&36593&-3.4 \\ 
23651&5100&0.0075&0.004&36749&-3.7 \\ 
23720&4400&0.0175&0.008&36934&-3.4 \\ 
23775&5100&0.0128&0.008&36886&-3.4 \\ 
23880&5100&0.0211&0.008&36733&-3.4 \\ 
23949&5100&0.0064&0.003&36455&-3.8 \\ 
24798&5100&0.0108&0.006&36721&-3.5 \\ 
25558&5100&0.0071&0.003&36902&-3.8 \\ 
25894&5100&0.0057&0.003&36593&-3.8 \\ 
25937&5100&0.0065&0.005&36871&-3.6 \\ 
26107&5100&0.0076&0.005&36554&-3.6 \\ 
26480&5100&0.0152&0.006&37122&-3.5 \\ 
26559&5100&0.0081&0.003&36833&-3.8 \\ 
26892&5100&0.0084&0.004&36558&-3.7 \\ 
26895&5100&0.0071&0.004&36501&-3.7 \\ 
26985&5100&0.0131&0.003&36871&-3.8 \\ 
27169&5100&0.0041&0.003&36606&-3.8 \\ 
27399&5100&0.0069&0.003&36437&-3.9 \\ 
27683&5100&0.0097&0.004&36954&-3.7 \\ 
27712&5100&0.0157&0.005&36613&-3.6 \\ 
27780&5100&0.0088&0.005&36502&-3.6 \\ 
27820&5100&0.0098&0.005&36887&-3.6 \\ 
28161&5100&0.0129&0.006&36770&-3.5 \\ 
28786&5100&0.0064&0.003&36605&-3.8 \\ 
29272&5100&0.0171&0.006&36735&-3.5 \\ 
29349&5100&0.007&0.003&36596&-3.8 \\ 
29398&5100&0.0091&0.005&36652&-3.6 \\ 
29640&216&0.1001&0.048&37642&-2.6 \\ 
31800&5100&0.007&0.003&36596&-3.8 \\ 
32019&5100&0.0048&0.004&36632&-3.7 \\ 
34941&5100&0.007&0.004&36642&-3.7 \\ 
35812&5100&0.0048&0.003&36581&-3.8 \\ 
36744&5100&0.0168&0.006&36734&-3.5 \\ 
37150&5100&0.0128&0.004&36672&-3.7 \\ 
37207&5100&0.0088&0.004&36624&-3.7 \\ 
37234&5100&0.0088&0.005&36656&-3.6 \\ 
37256&400&0.1437&0.049&37607&-2.6 \\ 
37265&5100&0.0086&0.003&36587&-3.8 \\ 
37384&1300&0.0229&0.01&36907&-3.3 \\ 
37677&5100&0.0084&0.003&36632&-3.8 \\ 
37776&5100&0.0059&0.004&36618&-3.7 \\ 
37933&5100&0.0105&0.005&36695&-3.6 \\ 
38049&5100&0.0072&0.003&36505&-3.8 \\ 
38331&5100&0.0052&0.004&36656&-3.7 \\ 
38694&3648&0.0191&0.007&35951&-3.5 \\ 
39256&5100&0.0068&0.003&36618&-3.8 \\ 
40141&5100&0.0045&0.003&36603&-3.8 \\ 
40547&1200&0.0623&0.027&37302&-2.8 \\ 
40872&5100&0.0094&0.003&35996&-3.8 \\ 
40896&5100&0.0085&0.004&37251&-3.7 \\ 
40990&5100&0.0081&0.003&36096&-3.8 \\ 
41021&4488&0.0192&0.008&36740&-3.4 \\ 
41034&5100&0.0145&0.006&36742&-3.5 \\ 
41194&5100&0.0116&0.005&36694&-3.6 \\ 
41241&1400&0.0202&0.008&36129&-3.4 \\ 
41380&5100&0.0102&0.004&36646&-3.7 \\ 
41434&100&0.0904&0.062&36277&-2.5 \\ 
41469&3472&0.028&0.009&36770&-3.4 \\ 
41586&5100&0.0063&0.003&36541&-3.8 \\ 
41745&5100&0.0074&0.004&36688&-3.7 \\ 
41838&5100&0.0046&0.003&36554&-3.8 \\ 
41882&5100&0.0096&0.006&36715&-3.5 \\ 
41903&5100&0.0112&0.003&36623&-3.8 \\ 
41911&5100&0.0077&0.004&36629&-3.7 \\ 
41944&5100&0.0097&0.003&36595&-3.8 \\ 
42662&5100&0.008&0.003&36630&-3.8 \\ 
42691&5100&0.0063&0.003&36590&-3.8 \\ 
42917&5100&0.0133&0.005&36711&-3.6 \\ 
42942&5100&0.0079&0.004&36629&-3.7 \\ 
42951&5100&0.0052&0.004&36618&-3.7 \\ 
42984&5100&0.0067&0.003&36596&-3.8 \\ 
43432&5100&0.0174&0.006&36726&-3.5 \\ 
43463&5100&0.0049&0.003&36601&-3.8 \\ 
43539&1448&0.0103&0.007&36529&-3.4 \\ 
43587&5100&0.011&0.004&36651&-3.7 \\ 
43823&5100&0.0157&0.006&36735&-3.5 \\ 
43874&5100&0.009&0.003&36615&-3.8 \\ 
44204&1500&0.019&0.012&36751&-3.2 \\ 
44337&1000&0.0201&0.011&36915&-3.2 \\ 
44709&400&0.05&0.023&36420&-2.9 \\ 
44910&5100&0.0041&0.004&36712&-3.7 \\ 
45246&5100&0.0094&0.006&36741&-3.5 \\ 
45807&5100&0.0084&0.004&36733&-3.7 \\ 
45920&5100&0.0087&0.003&36588&-3.8 \\ 
46112&5100&0.0067&0.004&36624&-3.7 \\

\end{longtable}
\end{center}

\begin{center}
\begin{longtable}{|c|c|l|l|l|l|l|}
\caption{Upper limits at 154 MHz} \label{tab4} \\

\hline \multicolumn{1}{|c|}{\textbf{NORAD ID \#}} & \multicolumn{1}{c|}{\textbf{Integration time (s)}} & \multicolumn{1}{c|}{\textbf{Pixel intensity (Jy/beam)}} & \multicolumn{1}{c|}{\textbf{RMS (Jy/beam)}} & \multicolumn{1}{c|}{\textbf{Mean distance (km)}} & \multicolumn{1}{c|}{\textbf{$Log_{10} (EIRP (W))$}}\\ \hline 
\endfirsthead

\multicolumn{3}{c}%
{{\bfseries \tablename\ \thetable{} -- continued from previous page}} \\
\hline \multicolumn{1}{|c|}{\textbf{NORAD ID \#}} & \multicolumn{1}{c|}{\textbf{Integration time (s)}} & \multicolumn{1}{c|}{\textbf{Pixel intensity (Jy/beam)}} & \multicolumn{1}{c|}{\textbf{RMS (Jy/beam)}} & \multicolumn{1}{c|}{\textbf{Mean distance (km)}} & \multicolumn{1}{c|}{\textbf{$Log_{10} (EIRP (W))$}}\\ \hline 
\endhead

\hline \multicolumn{3}{|r|}{{Continued on next page}} \\ \hline
\endfoot

\hline \hline
\endlastfoot
2217&796&0.0189&0.008&34495&-3.4 \\ 
5854&5000&0.0051&0.002&36512&-3.9 \\ 
8357&5000&0.0076&0.003&36927&-3.8 \\ 
8476&2464&0.0109&0.004&37146&-3.7 \\ 
10516&2600&0.0093&0.004&37281&-3.7 \\ 
11484&2888&0.0139&0.005&36812&-3.6 \\ 
11567&3468&0.0107&0.005&36760&-3.6 \\ 
11570&3400&0.0086&0.004&37710&-3.7 \\ 
11571&5000&0.0106&0.004&36726&-3.7 \\ 
12635&1800&0.0116&0.005&37278&-3.6 \\ 
13086&2500&0.0087&0.003&37398&-3.8 \\ 
14786&424&0.0446&0.027&38208&-2.8 \\ 
14948&2000&0.0164&0.005&36970&-3.6 \\ 
15236&1800&0.0176&0.006&37900&-3.5 \\ 
15545&2700&0.0134&0.005&36638&-3.6 \\ 
15826&3100&0.0069&0.003&36674&-3.8 \\ 
18570&3100&0.0067&0.003&37017&-3.8 \\ 
19090&912&0.0318&0.01&37682&-3.3 \\ 
19710&3256&0.0065&0.003&36555&-3.8 \\ 
19928&3276&0.0093&0.005&36705&-3.6 \\ 
20705&3300&0.0036&0.002&36992&-3.9 \\ 
21132&3200&0.0065&0.003&36536&-3.8 \\ 
21762&3100&0.0091&0.004&36850&-3.7 \\ 
21925&3100&0.0085&0.003&36357&-3.8 \\ 
22883&3300&0.004&0.003&36435&-3.9 \\ 
23111&3256&0.0065&0.004&35784&-3.7 \\ 
23651&5000&0.0052&0.002&36757&-3.9 \\ 
23720&3080&0.0124&0.005&36749&-3.6 \\ 
23775&2768&0.0071&0.004&36592&-3.7 \\ 
23880&780&0.0251&0.01&36193&-3.3 \\ 
23949&5000&0.005&0.002&36461&-4.0 \\ 
24798&4084&0.0119&0.005&36719&-3.6 \\ 
25558&5000&0.004&0.002&36904&-4.0 \\ 
25894&5000&0.0036&0.002&36593&-4.0 \\ 
26480&2116&0.0112&0.005&36769&-3.6 \\ 
26559&5000&0.0032&0.002&36833&-4.1 \\ 
26892&5000&0.0067&0.003&36567&-3.9 \\ 
26895&4800&0.0044&0.002&36482&-3.9 \\ 
26985&5000&0.0047&0.002&36871&-4.0 \\ 
27169&5000&0.0033&0.002&36609&-4.0 \\ 
27399&5000&0.0041&0.002&36438&-4.1 \\ 
27683&5000&0.0113&0.003&36953&-3.9 \\ 
27712&5000&0.0104&0.004&36622&-3.7 \\ 
27780&5000&0.0079&0.004&36509&-3.7 \\ 
27820&240&0.047&0.023&36936&-2.9 \\ 
28161&1008&0.0286&0.016&37233&-3.1 \\ 
28786&5000&0.0056&0.002&36605&-4.0 \\ 
29349&5000&0.0026&0.002&36596&-4.0 \\ 
31800&5000&0.0036&0.002&36596&-4.0 \\ 
32019&5000&0.0034&0.002&36632&-3.9 \\ 
34941&5000&0.0075&0.003&36642&-3.8 \\ 
35812&5000&0.004&0.002&36581&-4.0 \\ 
37150&5000&0.0091&0.004&36672&-3.7 \\ 
37207&5000&0.0035&0.002&36624&-3.9 \\ 
37234&1472&0.0173&0.007&36660&-3.4 \\ 
37256&136&0.1009&0.037&37448&-2.7 \\ 
37265&5000&0.0044&0.002&36587&-4.0 \\ 
37384&1000&0.0162&0.008&36809&-3.4 \\ 
37677&5000&0.0038&0.002&36632&-3.9 \\ 
37776&5000&0.0047&0.003&36618&-3.9 \\ 
38049&5000&0.0033&0.002&36506&-4.1 \\ 
38331&5000&0.0053&0.003&36656&-3.7 \\ 
38694&424&0.0462&0.015&35602&-3.1 \\ 
39256&5000&0.0071&0.002&36618&-4.0 \\ 
40141&5000&0.004&0.002&36603&-3.9 \\ 
40547&400&0.0379&0.013&36973&-3.2 \\ 
40872&5000&0.0046&0.002&36003&-4.0 \\ 
40896&2764&0.0116&0.003&37275&-3.7 \\ 
40990&5000&0.0052&0.002&36100&-4.0 \\ 
41241&1000&0.0125&0.006&36192&-3.5 \\ 
41380&5000&0.0049&0.003&36646&-3.8 \\ 
41586&5000&0.0039&0.002&36541&-4.0 \\ 
41745&5000&0.0084&0.004&36689&-3.7 \\ 
41838&5000&0.0041&0.002&36554&-4.1 \\ 
41903&5000&0.0035&0.002&36623&-3.9 \\ 
41911&5000&0.0053&0.003&36629&-3.8 \\ 
41944&5000&0.0043&0.002&36595&-4.1 \\ 
42662&5000&0.005&0.002&36630&-3.9 \\ 
42691&5000&0.0032&0.002&36590&-4.0 \\ 
42942&5000&0.0066&0.003&36629&-3.8 \\ 
42951&5000&0.0047&0.002&36618&-3.9 \\ 
42984&5000&0.0023&0.002&36596&-4.0 \\ 
43463&5000&0.0024&0.002&36601&-4.0 \\ 
43539&1100&0.0102&0.005&36441&-3.6 \\ 
43587&5000&0.0061&0.003&36651&-3.8 \\ 
43874&5000&0.0055&0.002&36616&-3.9 \\ 
44204&1092&0.0186&0.01&36714&-3.3 \\ 
44337&500&0.0417&0.011&36921&-3.2 \\ 
44709&400&0.0332&0.014&36443&-3.1 \\ 
44910&5000&0.0053&0.003&36713&-3.8 \\ 
45807&5000&0.005&0.002&36734&-3.9 \\ 
45920&5000&0.0031&0.002&36588&-4.0 \\ 
46112&5000&0.0041&0.002&36624&-3.9 \\

\end{longtable}
\end{center}

\begin{center}
\begin{longtable}{|c|c|l|l|l|l|l|}
\caption{Upper limits at 185 MHz} \label{tab5} \\

\hline \multicolumn{1}{|c|}{\textbf{NORAD ID \#}} & \multicolumn{1}{c|}{\textbf{Integration time (s)}} & \multicolumn{1}{c|}{\textbf{Pixel intensity (Jy/beam)}} & \multicolumn{1}{c|}{\textbf{RMS (Jy/beam)}} & \multicolumn{1}{c|}{\textbf{Mean distance (km)}} & \multicolumn{1}{c|}{\textbf{$Log_{10} (EIRP (W))$}}\\ \hline 
\endfirsthead

\multicolumn{3}{c}%
{{\bfseries \tablename\ \thetable{} -- continued from previous page}} \\
\hline \multicolumn{1}{|c|}{\textbf{NORAD ID \#}} & \multicolumn{1}{c|}{\textbf{Integration time (s)}} & \multicolumn{1}{c|}{\textbf{Pixel intensity (Jy/beam)}} & \multicolumn{1}{c|}{\textbf{RMS (Jy/beam)}} & \multicolumn{1}{c|}{\textbf{Mean distance (km)}} & \multicolumn{1}{c|}{\textbf{$Log_{10} (EIRP (W))$}}\\ \hline 
\endhead

\hline \multicolumn{3}{|r|}{{Continued on next page}} \\ \hline
\endfoot

\hline \hline
\endlastfoot
2217&300&0.0474&0.016&34458&-3.1 \\ 
5854&5100&0.0055&0.003&36513&-3.9 \\ 
8357&5100&0.0106&0.004&36924&-3.7 \\ 
8476&1800&0.0113&0.004&37021&-3.6 \\ 
10516&1876&0.008&0.004&37148&-3.7 \\ 
11484&88&0.0983&0.039&36313&-2.7 \\ 
11570&1800&0.0154&0.005&37572&-3.5 \\ 
11571&3200&0.009&0.006&36650&-3.5 \\ 
12635&1500&0.0094&0.005&37175&-3.5 \\ 
13086&2096&0.0091&0.004&37300&-3.7 \\ 
14948&1600&0.0116&0.006&36880&-3.5 \\ 
15236&728&0.0396&0.015&37942&-3.1 \\ 
15545&1308&0.0115&0.006&36261&-3.5 \\ 
15826&2700&0.0092&0.003&36680&-3.8 \\ 
18570&2800&0.0096&0.004&36959&-3.7 \\ 
19710&2800&0.0039&0.003&36500&-3.9 \\ 
19928&1532&0.0119&0.006&36386&-3.5 \\ 
20705&2900&0.0038&0.003&36900&-3.8 \\ 
21132&2776&0.0104&0.004&36568&-3.7 \\ 
21762&2532&0.0174&0.005&36814&-3.6 \\ 
21925&2800&0.0084&0.004&36277&-3.7 \\ 
22883&2800&0.0077&0.002&36447&-3.9 \\ 
23651&4932&0.0059&0.003&36730&-3.8 \\ 
23720&1016&0.0313&0.012&37004&-3.2 \\ 
23775&512&0.0214&0.012&36348&-3.2 \\ 
23949&5100&0.0049&0.002&36458&-4.0 \\ 
25558&5100&0.0038&0.002&36903&-4.0 \\ 
25894&5100&0.0041&0.002&36593&-4.1 \\ 
26559&5100&0.0033&0.001&36833&-4.1 \\ 
26892&4168&0.0074&0.003&36499&-3.9 \\ 
26895&4200&0.0081&0.002&36380&-3.9 \\ 
26985&5100&0.0046&0.002&36872&-3.9 \\ 
27169&5100&0.004&0.002&36607&-3.9 \\ 
27399&5100&0.0032&0.001&36438&-4.1 \\ 
27683&3984&0.0064&0.003&36951&-3.8 \\ 
28786&5100&0.0047&0.002&36605&-3.9 \\ 
29349&5100&0.004&0.002&36596&-3.9 \\ 
31800&5100&0.0059&0.002&36596&-4.0 \\ 
32019&5100&0.0079&0.003&36632&-3.8 \\ 
34941&5100&0.0063&0.004&36642&-3.7 \\ 
35812&5100&0.0025&0.002&36581&-4.0 \\ 
37207&5100&0.0082&0.004&36624&-3.7 \\ 
37265&5100&0.0034&0.002&36587&-4.0 \\ 
37384&700&0.0153&0.008&36745&-3.4 \\ 
37677&5100&0.0052&0.003&36632&-3.9 \\ 
37776&5100&0.0099&0.004&36618&-3.7 \\ 
38049&5100&0.0035&0.002&36506&-4.1 \\ 
38331&5100&0.0091&0.004&36656&-3.6 \\ 
39256&5100&0.0046&0.002&36618&-4.0 \\ 
40141&5100&0.0065&0.003&36603&-3.8 \\ 
40547&288&0.0509&0.016&36910&-3.1 \\ 
40872&5100&0.004&0.002&35998&-4.0 \\ 
40896&816&0.0211&0.008&37237&-3.4 \\ 
40990&5100&0.0034&0.002&36097&-4.0 \\ 
41241&600&0.0116&0.005&36294&-3.6 \\ 
41380&5100&0.0098&0.004&36646&-3.6 \\ 
41586&5100&0.005&0.002&36541&-3.9 \\ 
41838&5100&0.0025&0.002&36554&-4.0 \\ 
41903&5100&0.0062&0.003&36623&-3.8 \\ 
41944&5100&0.0044&0.002&36595&-4.0 \\ 
42662&5100&0.0064&0.002&36630&-3.9 \\ 
42691&5100&0.0048&0.002&36590&-4.0 \\ 
42942&5100&0.006&0.003&36629&-3.8 \\ 
42951&5100&0.0049&0.004&36618&-3.7 \\ 
42984&5100&0.0044&0.002&36596&-3.9 \\ 
43463&5100&0.0059&0.002&36601&-3.9 \\ 
43539&900&0.0102&0.005&36402&-3.6 \\ 
43587&2184&0.0112&0.006&36651&-3.5 \\ 
43874&5100&0.011&0.003&36616&-3.8 \\ 
44204&800&0.0159&0.008&36661&-3.4 \\ 
44337&200&0.0489&0.024&36950&-2.9 \\ 
44709&200&0.0472&0.018&36346&-3.0 \\ 
44910&1388&0.0132&0.007&36761&-3.5 \\ 
45807&5100&0.0061&0.003&36733&-3.9 \\ 
45920&5100&0.0067&0.002&36588&-3.9 \\ 
46112&5100&0.0094&0.004&36624&-3.7 \\

\end{longtable}
\end{center}

\begin{center}
\begin{longtable}{|c|c|l|l|l|l|l|}
\caption{Upper limits at 216 MHz}  \label{tab6} \\

\hline \multicolumn{1}{|c|}{\textbf{NORAD ID \#}} & \multicolumn{1}{c|}{\textbf{Integration time (s)}} & \multicolumn{1}{c|}{\textbf{Pixel intensity (Jy/beam)}} & \multicolumn{1}{c|}{\textbf{RMS (Jy/beam)}} & \multicolumn{1}{c|}{\textbf{Mean distance (km)}} & \multicolumn{1}{c|}{\textbf{$Log_{10} (EIRP (W))$}}\\ \hline 
\endfirsthead

\multicolumn{3}{c}%
{{\bfseries \tablename\ \thetable{} -- continued from previous page}} \\
\hline \multicolumn{1}{|c|}{\textbf{NORAD ID \#}} & \multicolumn{1}{c|}{\textbf{Integration time (s)}} & \multicolumn{1}{c|}{\textbf{Pixel intensity (Jy/beam)}} & \multicolumn{1}{c|}{\textbf{RMS (Jy/beam)}} & \multicolumn{1}{c|}{\textbf{Mean distance (km)}} & \multicolumn{1}{c|}{\textbf{$Log_{10} (EIRP (W))$}}\\ \hline 
\endhead

\hline \multicolumn{3}{|r|}{{Continued on next page}} \\ \hline
\endfoot

\hline \hline
\endlastfoot
5854&5100&0.0123&0.005&36513&-3.6 \\ 
8357&2580&0.0171&0.008&36838&-3.4 \\ 
8476&1400&0.0211&0.007&36952&-3.4 \\ 
10516&1500&0.0135&0.006&37087&-3.5 \\ 
11570&1100&0.0192&0.008&37482&-3.4 \\ 
11571&1980&0.0551&0.017&36587&-3.1 \\ 
12635&1272&0.0157&0.009&37138&-3.3 \\ 
13086&1700&0.0074&0.005&37230&-3.6 \\ 
14948&1296&0.0154&0.01&36831&-3.3 \\ 
15545&8&0.3336&0.138&36006&-2.2 \\ 
15826&2392&0.0126&0.005&36665&-3.6 \\ 
18570&2500&0.0116&0.005&36924&-3.6 \\ 
19710&2600&0.0097&0.004&36503&-3.7 \\ 
20705&2500&0.0098&0.004&36934&-3.7 \\ 
21132&2500&0.0143&0.007&36522&-3.4 \\ 
21762&960&0.0177&0.013&36993&-3.2 \\ 
21925&2300&0.0149&0.007&36260&-3.5 \\ 
22883&2556&0.0058&0.003&36409&-3.8 \\ 
23651&3800&0.0159&0.005&36727&-3.6 \\ 
23949&4900&0.0075&0.004&36440&-3.7 \\ 
25558&5100&0.0063&0.004&36905&-3.7 \\ 
25894&5100&0.0067&0.003&36593&-3.9 \\ 
26559&5100&0.0073&0.002&36833&-4.0 \\ 
26892&3240&0.0072&0.004&36487&-3.7 \\ 
26895&3276&0.0068&0.003&36365&-3.8 \\ 
26985&5064&0.0078&0.004&36871&-3.7 \\ 
27169&5000&0.0121&0.005&36604&-3.6 \\ 
27399&5100&0.0041&0.002&36438&-4.0 \\ 
27683&324&0.0532&0.031&36953&-2.8 \\ 
28786&5100&0.0063&0.003&36605&-3.8 \\ 
29349&5100&0.0157&0.006&36596&-3.5 \\ 
31800&5100&0.0057&0.003&36596&-3.8 \\ 
32019&5100&0.0115&0.006&36632&-3.5 \\ 
35812&5100&0.0095&0.003&36581&-3.8 \\ 
37207&5100&0.0223&0.011&36624&-3.3 \\ 
37265&5100&0.0069&0.003&36587&-3.8 \\ 
37384&524&0.0315&0.011&36775&-3.2 \\ 
37677&5100&0.0111&0.005&36632&-3.6 \\ 
37776&5100&0.0194&0.012&36618&-3.2 \\ 
38049&5100&0.0032&0.003&36506&-3.9 \\ 
39256&5100&0.0105&0.004&36618&-3.7 \\ 
40141&5100&0.0081&0.004&36603&-3.7 \\ 
40547&144&0.1021&0.039&36840&-2.7 \\ 
40872&5100&0.0086&0.004&36001&-3.7 \\ 
40990&5100&0.0063&0.003&36098&-3.8 \\ 
41241&500&0.0227&0.011&36330&-3.3 \\ 
41586&5100&0.0113&0.003&36542&-3.8 \\ 
41838&5100&0.0088&0.003&36554&-3.9 \\ 
41903&5100&0.0194&0.008&36623&-3.4 \\ 
41944&5100&0.0077&0.003&36595&-3.8 \\ 
42662&5100&0.012&0.004&36630&-3.6 \\ 
42691&5100&0.0079&0.003&36590&-3.8 \\ 
42942&5100&0.0107&0.005&36629&-3.6 \\ 
42951&5100&0.0168&0.011&36618&-3.2 \\ 
42984&5100&0.0123&0.006&36596&-3.6 \\ 
43463&5100&0.0085&0.003&36601&-3.8 \\ 
43539&700&0.0107&0.008&36476&-3.4 \\ 
43874&5100&0.0207&0.009&36616&-3.3 \\ 
44204&700&0.0193&0.013&36616&-3.2 \\ 
44709&100&0.0925&0.035&36313&-2.8 \\ 
45807&5100&0.0084&0.004&36733&-3.7 \\ 
45920&5100&0.0059&0.003&36588&-3.8 \\ 
46112&5100&0.0181&0.009&36624&-3.3 \\

\end{longtable}
\end{center}

\end{document}